\documentclass[12pt]{iopart}

\usepackage{color}
\usepackage{graphicx}
\usepackage{subfig}
\usepackage{amssymb}
\expandafter\let\csname equation*\endcsname\relax 
\expandafter\let\csname endequation*\endcsname\relax 
\usepackage{amsmath}
\usepackage{cite}
\graphicspath{ {pics/} }
\newcommand{\beginsupplement}{%
        \setcounter{table}{0}
        \renewcommand{\thetable}{S\arabic{table}}%
        \setcounter{figure}{0}
        \renewcommand{\thefigure}{S\arabic{figure}}%
        \setcounter{equation}{0}
        \renewcommand{\theequation}{S\arabic{equation}}%
        \setcounter{section}{0}
        \renewcommand{\thesection}{S\arabic{section}}%
     }
\begin{document}

\title[Resolution limit revisited.]{Resolution limit revisited: community detection using generalized modularity density.}

\author{Jiahao Guo\textsuperscript{1,2}, Pramesh Singh\textsuperscript{1,2}, Kevin E. Bassler\textsuperscript{1,2,3}}
\address{\textsuperscript{1}~Department of Physics, University of Houston, Houston, Texas 77204, USA.\\
\textsuperscript{2}~Texas Center for Superconductivity, University of Houston, Houston 77204, Texas, USA.\\
\textsuperscript{3}~Department of Mathematics, University of Houston, Houston, Texas 77204, USA.}
\ead{bassler@uh.edu}


\begin{abstract}
Various attempts have been made in recent years to solve the Resolution Limit (RL) problem
in community detection by considering variants of the modularity metric in the detection algorithms. These
metrics purportedly largely mitigate the RL problem and are preferable to modularity in many realistic
scenarios. However, they are not generally suitable for analyzing weighted networks or for detecting
hierarchical community structure. Resolution limit problems can be complicated, though, and in particular it can
be unclear when it should be considered as problem. In this paper, we introduce a metric that we call
generalized modularity density $Q_g$ that eliminates the RL problem at any desired resolution and is easily
extendable to study weighted and hierarchical networks. We also propose a benchmark test to
quantify the resolution limit problem, examine various modularity-like metrics to show that the new metric $Q_g$
performs best, and show that $Q_g$ can identify modular structure in real-world and artificial networks that is
otherwise hidden.

\end{abstract}

\maketitle

\section{Introduction}
Networks are excellent tools for describing complex biological, social, and infrastructural systems~\cite{newmanbook,chauhan2016,trevino2012,Bhavnani2012}. 
Most real-world examples of complex networks are far from being random and have a community or modular structure within them~\cite{newman2003,danon2005,fortunato2010}. Detecting this structure is crucial in understanding the function and dynamics of a complex network. Although there is no universally accepted definition of a community structure~\cite{schaub2017,peel2017}, it is often characterized by dense connectivity within groups and sparser connectivity between different groups. Modularity,  $Q$, is a widely used metric to quantify the presence of this type of structure~\cite{newman2003,newman2002,newman2004,newman2004epj,sun2009,trevino2015,guo2019}. For a partition of the nodes of an unweighted network, $C = \{c_1, c_2, c_3, ..\}$, it is defined as 
\begin{equation} 
Q = \frac{1}{2m}\sum_{c \in C}\left(2m_c -\frac{K_c^2}{2m}\right)
\label{Q}
\end{equation}
where $m_c$ is the number of links in community $c$,
$K_c$ is the sum of degrees of nodes in $c$,
and $m$ is the total number of links in the network. $Q$ measures the difference between the fraction of links within communities and the expected fraction if the links were randomly placed. The partition that maximizes the metric $Q$ identifies the community structure of the network. 
Despite its intuitively appealing definition, there is a fundamental problem with using $Q$ to find community structure. Namely, communities smaller than a certain size in large network may not be detected. 
This {\it Resolution Limit} (RL) problem~\cite{fortunato2007,traag2011} reduces the domain of applicability of $Q$ and
is often a significant issue when analyzing empirical networks.

Alternate metrics have been proposed in recent years~\cite{ronhovde2010,arenas2008,granell2012,aldecova2011,mingming2013,mingming2014,charo2016,chen2018,haq2019} to mitigate the RL problem. Some of these metrics~\cite{mingming2013,mingming2014,charo2016,chen2018,haq2019}, known as modularity density metrics, weights that are functions of the internal link density of communities are applied to the two terms in Eq.~\ref{Q}.
In this paper we propose a new metric of this form, which we call {\it generalized modularity density} $Q_g$. $Q_g$ is an extension of $Q$, as it reduces to $Q$ in a limit. The main reasons for introducing this new metric are as follows. 
First, it has an adjustable parameter $\chi$ that controls the resolution density of the communities that are detected.
Second, $Q_g$ can be extended to detect communities in weighted networks in a way that has a clear interpretation and is independent of the scale of the link weights.

The RL problem can be seen in the simple example of cliques arranged in a ring connected to one another in series by single links~\cite{fortunato2007}. The expectation in this case is that the cliques should be detected as separate communities. Unfortunately, with some metrics, pairs of cliques are merged into the same community. Of course, if all possible cross-links between two cliques are present, then it is sensible to merge them into one community as they simply form a clique of larger size.
However, when cliques are connected by an intermediate number of links or when the network is weighted, it is unclear whether the cliques should be merged or separated~\cite{lancichinetti2011}.
Intuitively, it makes sense to merge two cliques at sufficiently high density of cross-links. Generally, methods of community detection that use different metrics have a different critical value for this density. The answer may also depend on the specific application being considered. Thus, it is useful to have some flexibility in allowing the communities to be separated or merged. $Q_g$ achieves this goal by varying a parameter $\chi$. We will show that for a properly chosen value of $\chi$, the partition that maximizes $Q_g$ separates two cliques at any desired strength of inter-connectivity. This tunability of our metric is extremely useful for analyzing networks that exhibit hierarchical community structure~\cite{newman2003}, which is found in many real-world networks. A common way to investigate these hierarchical structures is to iteratively perform community detection within detected communities~\cite{park2019global}. Using our approach, one can simply vary $\chi$.

Finally, we compare the performance of our metric against other modularity density metrics by using them to find the structure in a more complex benchmark network than a simple ring of cliques. Our analysis indicates that $Q_g$ performs better than all other metrics considered. We then use $Q_g$ to find structure in a variety of empirical and artificial networks to demonstrate its ability to detect hidden community structure. We find that it eliminates the resolution limit problem that we consider and that it is applicable to a wider range of problems than other metrics. In addition, the network partition that maximizes $Q_g$ can be efficiently and accurately found using the recently introduced Reduced Network Extremal Ensemble Learning (RenEEL) scheme~\cite{guo2019}.

\section{Methods}
\subsection{Generalized Modularity Density}
We define the \emph{Generalized Modularity Density} of a node partition of unweighted network as
\begin{equation}    \label{eq:Qg}
Q_g=\frac{1}{2m}\sum_c (2m_c-\frac{K_c^2}{2m})\rho_c^\chi
\end{equation}
where $m$ is the number of total links of the network, $m_c$ is the number of links within a community c, $K_c$ is the sum of degrees of all nodes in community c, $\rho_c$ is the link density of community c, the exponent $\chi$ is a control parameter. Here we assume that $\chi$ is a non-negative real number. The link density of a community is the ratio of the number of links that exist in $c$ to the number of possible links that can exist
\begin{equation}     \label{eq:rel-density}
     \rho_c=\frac{2 m_c}{n_c(n_c-1)} \; ,
\end{equation} where $n_c$ is the number of nodes in $c$. 
$Q_g$ is an extension of modularity, i.e. at $\chi = 0$, $Q_g=Q$.

The metric $Q_g$, like the Modularity metric $Q$ (Eq.~\ref{Q}), can be easily extended to weighted networks. For $Q$ this is done by simply replacing the number of links with the sum of link weights in $m$, $m_c$ and $K_c$~\cite{newman2004weighted,newman2008directed}.
Extending the definition of modularity density metrics to weighted networks is complicated by the fact that they depend on link density, and link density can be problematic to use with weighted networks.
One way to deal with these problems is to simply ignore the link weights and calculate the link density as if the network was unweighted~\cite{mingming2013,mingming2014}. Unfortunately, this loses the information contained in the link weights.
The correct way is to use a normalized definition of link density, where the sum of the weight of all internal links divided by the maximum value that sum would have if the community were fully connected with links of weight equal to the maximum weight of any link in the network,
\begin{equation} \label{eq:abs-density}
    \rho_c=\frac{2 m_c}{n_c(n_c-1)w_{\max}}
\end{equation}
where $m_c$ is the sum of the weights within community $c$, $n_c$ is the number of nodes in $c$, and $w_{\max}$ is the maximum weight of any link in the network.
This definition of $\rho_c$ is consistent with the definition for unweighted networks, 
but it can be problematic because it involves the global variable $w_{\max}$.
The community structure found using some metrics, such as those proposed in 
Refs.~\citeonline{mingming2013,chen2018,haq2019}, can be very sensitive to the value of $w_{\max}$. 
This makes their use potentially troublesome, especially in empirical studies where the value of $w_{\max}$ can be difficult to accurately measure. 
Additionally, if there is a wide distribution of link weights and $w_{\max}\rightarrow \infty$, then $\rho_c \rightarrow 0$ for all communities and the algorithms for finding the partition that maximizes the modularity density metric become numerically unstable. 

Generalized Modularity Density, unlike other modularity density metrics, does not have problems with $w_{\max}$. Both terms in Eq.~\ref{eq:Qg} are weighted by the same function of $w_{\max}$, which can factored out and simply modifies the value of $Q_g$ for every possible partition by the same constant factor. It is, thus, irrelevant for determining the partition that maximizes $Q_g$. So, instead of the absolute link density, Eq.~\ref{eq:abs-density}, a relative link density, given by Eq.~\ref{eq:rel-density} with $m_c$ being the sum of the weight of links in $c$,
can be used in the metric $Q_g$ without affecting results.
The community partitions found with Generalized Modularity Density are also independent of the scale of the link weights.  As it is with Modularity, multiplying all link weights by a common factor does not affect the results obtained with $Q_g$. This important property is needed for preserving the information in the link weights.

\subsection{Resolution Density}
\label{sec:resolutionscale}
The RL problem can be viewed as a problem with a metric, when using it yields a partition that merges two ``well separated" communities. 
A resolution-limit-free metric is expected to resolve these communities. 
Conversely, a metric should also avoid splitting two groups of nodes that are ``well connected" to each other. The RL problem is clear at these two extremes. However, more generally, the notion of well separated/connected communities is not well defined. 
It is unclear whether two partially connected communities should be merged or not.

\begin{figure}
    \centering
    \includegraphics[width=0.6\textwidth]{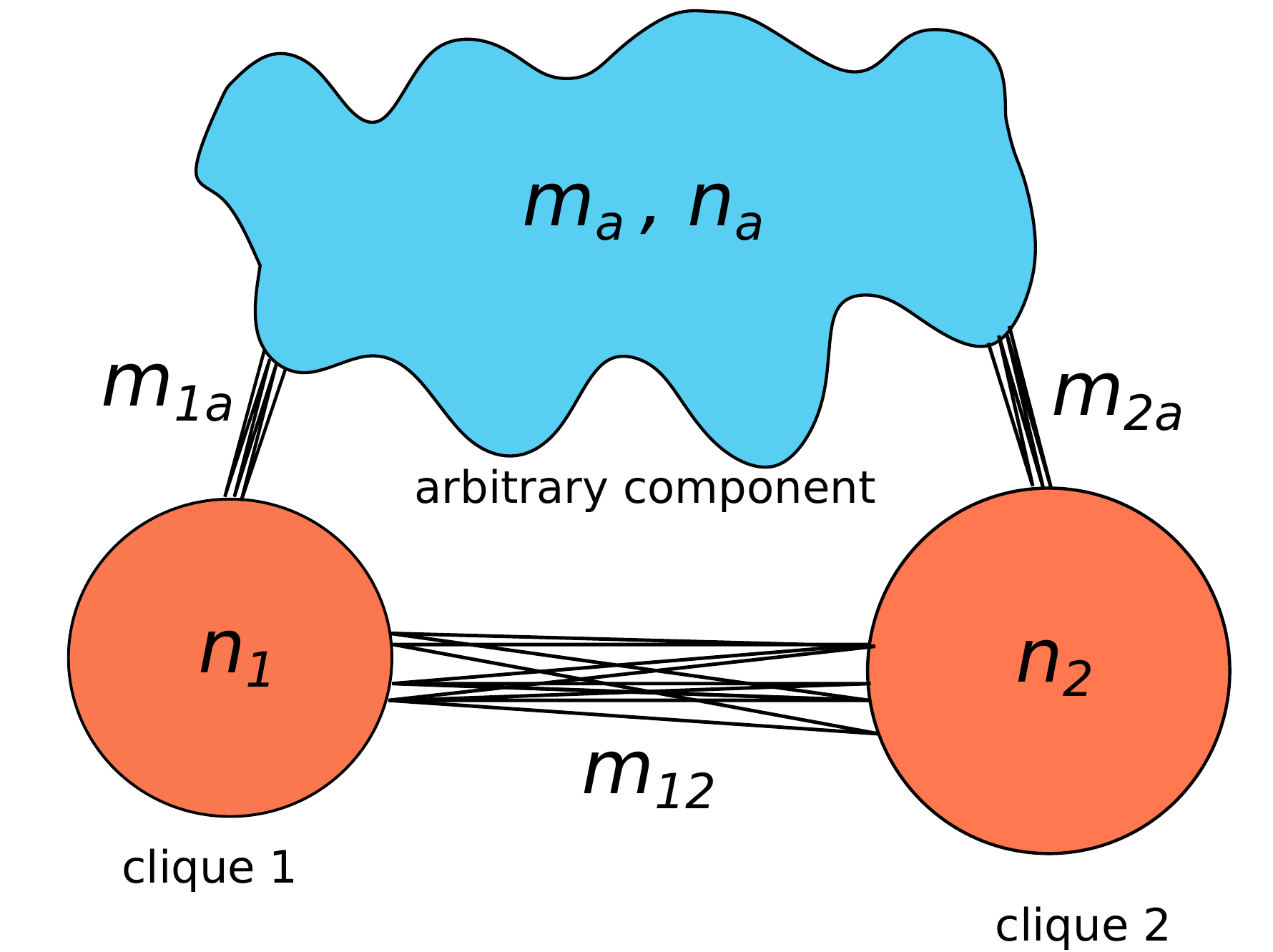}
    \caption{{\bf Benchmark network for studying the resolution limit problem.} The network consists of two cliques of sizes $n_1$ and $n_2$ and an arbitrary component with $n_a$ nodes and $m_a$ links. The two cliques share $m_{1a}$ and $m_{2a}$ links with the arbitrary component, respectively, and have $m_{12}$ links between. The links of the network can be weighted, in which case, $m_a$, $m_{1a}$, $m_{2a}$ and $m_{12}$ are the sums of link weights.}
    \label{fig:template}
\end{figure}
Consider the benchmark network shown in Fig.~\ref{fig:template}. This network consists of three parts: two cliques and an external arbitrary component to which the cliques are weakly connected.
As the cliques are fully connected, they have no internal community structure.
Assume clique 1 has $n_1$ nodes, clique 2 has $n_2$ nodes, and both $n_1$ and $n_2 \geq 3$. Without loss of generality, we assume $n_2 \ge n_1$. Let $m_{12}$ be the sum of weights of links between the two cliques, and let $m_{1a}$ and $m_{2a}$ be the sum of the weights of links that connect each clique with the arbitrary component. $n_a$ and $m_a$ are number of nodes and the sum of weights of links within the arbitrary component, respectively.
Without loss of generality, assume $n_1 \leq n_2$.
Also, assume that $m_{1a} \ll n_1^2w_{\max}$ and $m_{2a} \ll n_2^2w_{\max}$, so that the cliques are only weakly connected to the arbitrary component.
The RL question concerning this network is whether or not the two cliques should be merged or split, and whether or not using a given metric will meet this expectation.
This choice of network gives greater flexibility to explore the RL problem than a simple ring of cliques, since the external component can have an arbitrary structure and the strength of inter-connectivity between the two cliques can be varied. 
Generally, there is a threshold, or critical, value of $m_{12}$ below which the cliques are separated and above which they are merged.
We impose an arbitrary \emph{expected critical value} $m_{\rm exp}$ such that the cliques should be merged if $m_{12} \geq m_{\rm exp}$ and separated if $m_{12}<m_{\rm exp}$. 
Instead of using the values of $m_{12}$ and $m_{\rm exp}$, it is convenient to use normalized inter-clique link density 
\begin{equation}
d=\frac{m_{12}}{n_1n_2 w_{\max}}
\end{equation}
and normalized expected critical resolution link density 
\begin{equation}
\delta_{\rm exp} =\frac{m_{\rm exp}}{n_1n_2 w_{\max}} \; .
\end{equation}
For unweighted networks $w_{\max} = 1$.

Given a metric, we can examine the RL question in the benchmark network. For a given set of network parameters, the two cliques are either merged or split. 
Accordingly, the parameter space can be divided into Merged (M) and Split (S) phases.
The value of the link density at the boundary of the two phases is $\delta$.
At the same time, there is an expected result, corresponding a specific understanding of the problem, given by $\delta_{\rm exp}$. 
The metric can then be evaluated by comparing the results obtained by using it with the expected results. 
Specifically, we define a resolution-limit-free metric as one for which $\delta \geq \delta_{\rm exp}$ for all parameters of the benchmark network.
Then, the metric is resolution-limit-free with respect to the expected resolution density.

\section{Results}

\subsection{Benchmark Test}
\label{sec:benchmark}
We now analytically study the extent to which the RL exists in benchmark network of Fig.~\ref{fig:template} when $Q_g$ is used as the metric. We also compare the results to that obtained when using other metrics. 
Whether the use of the metric $Q_g$ will split the cliques or not is determined by the sign of $\Delta Q_g=Q_{g}^{merge}-Q_{g}^{split}$, where $Q_g^{merged}$ and $Q_g^{split}$ are the values of $Q_g$ if the cliques are merged or split, respectively.
Let us define the variables
\begin{equation}
    p=\frac{n_1}{n_2}
\end{equation}
and
\begin{equation}
    t=\frac{m_a}{n_1n_2 w_{\max}}\;.
\end{equation}
$p\in(0,1]$ is the ratio size of the cliques. $t\in[0,\infty)$ measures the external influence on them. Then, 
\begin{equation}
\label{eq:dQg}
    \Delta Q_g \sim \Big(\frac{r+2d}{r+2}\Big)^\chi\Big(2d+r-\frac{(r+2d)^2}{r+2d+2t}\Big)-\Big(r-\frac{r^2-2+2d^2+2dr}{r+2d+2t}\Big)
\end{equation}
where
$r=p+1/p$.
If $\Delta Q_g<0$, splitting is preferred, and if $\Delta Q_g>0$, merging is preferred. 
Eq.~\ref{eq:dQg} determines whether the use of the metric $Q_g$, for a given value of $\chi$, will lead to M or S phase as a function of the variables $(p,d,t)$. 
The value of $d$ at which the phase boundary separating the M and S phases occurs is $\delta_{Q_g}$.

\begin{figure}
\centering
(a)\includegraphics[width=0.45\textwidth]{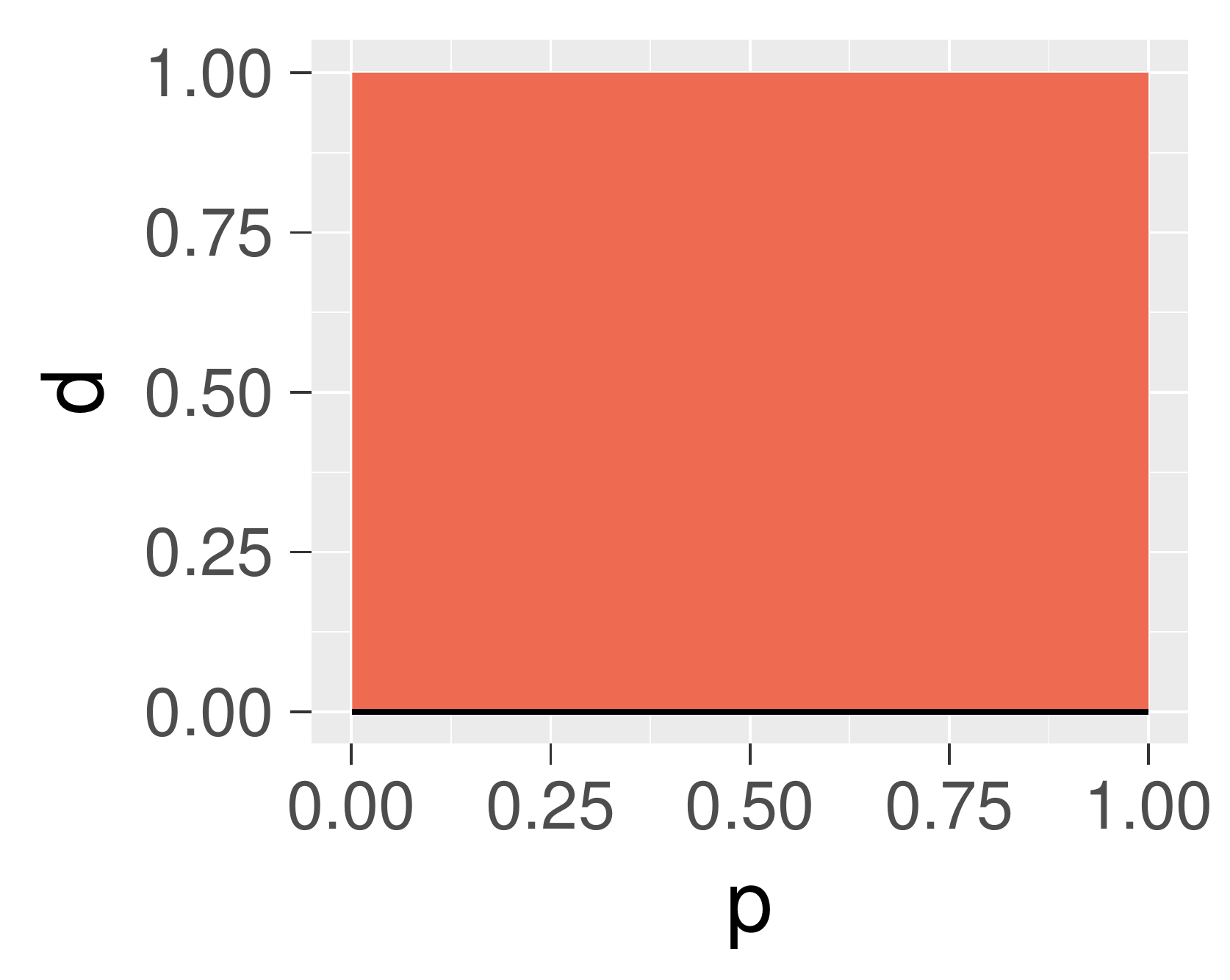}
(b)\includegraphics[width=0.45\textwidth]{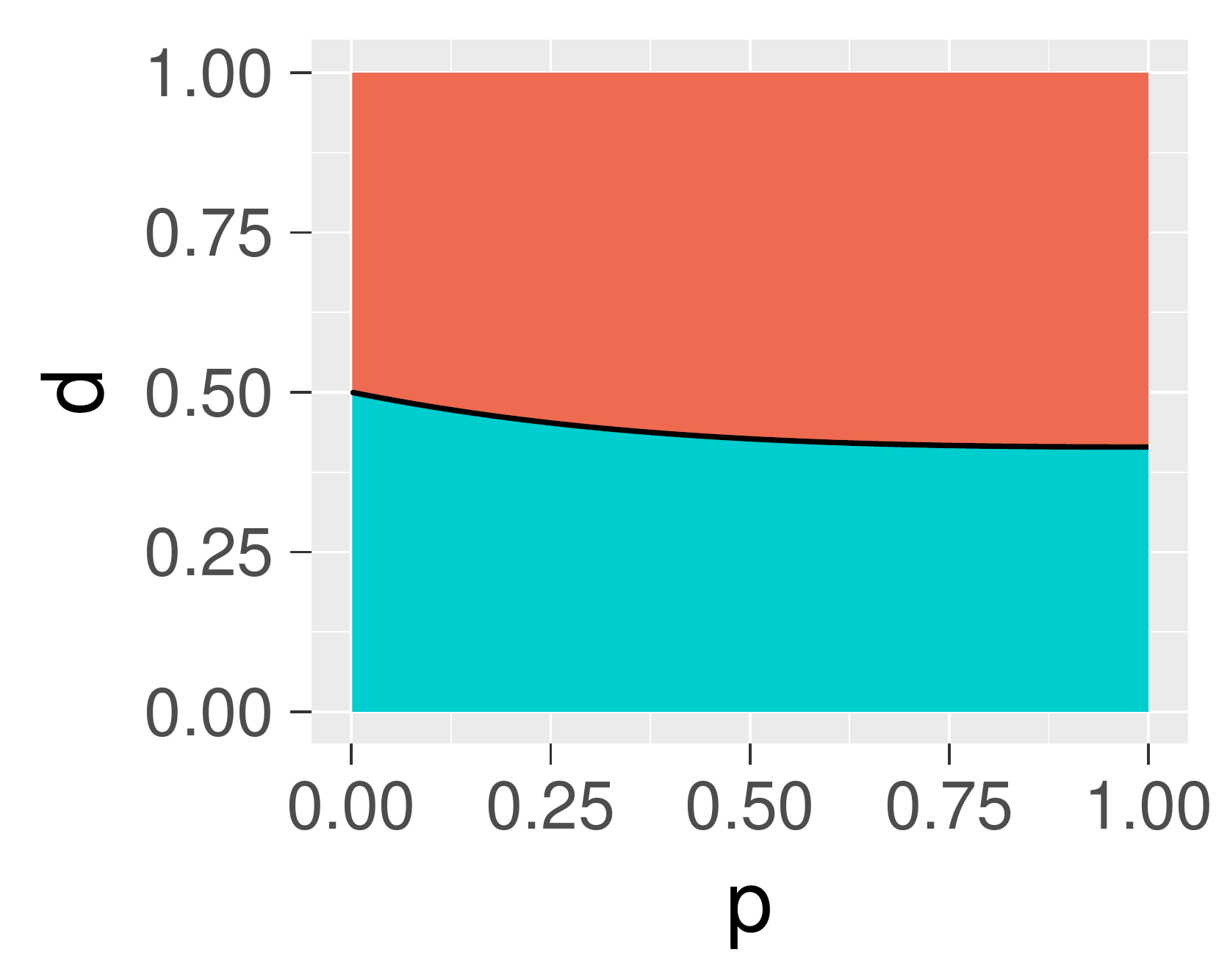}
(c)\includegraphics[width=0.45\textwidth]{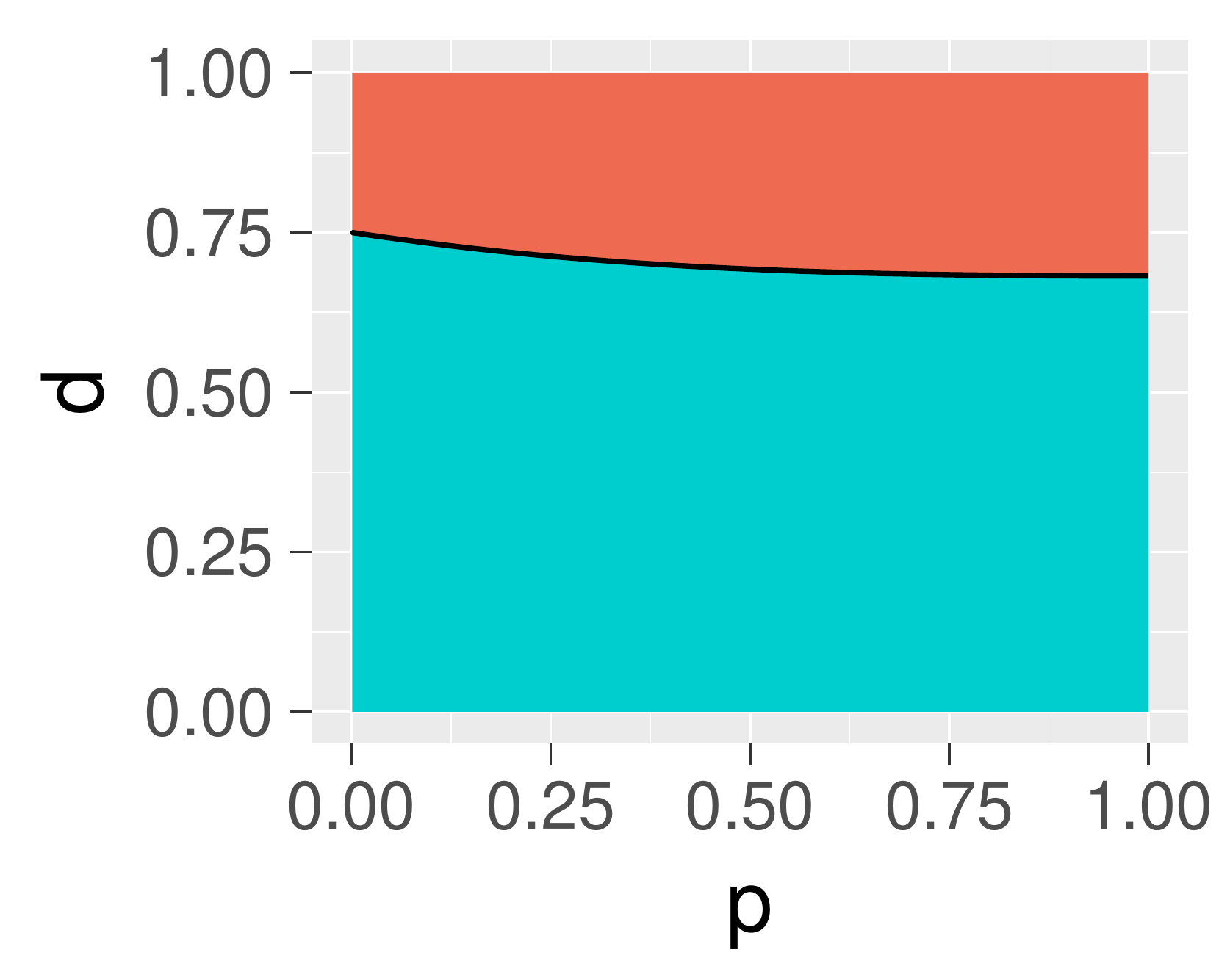}
(d)\includegraphics[width=0.45\textwidth]{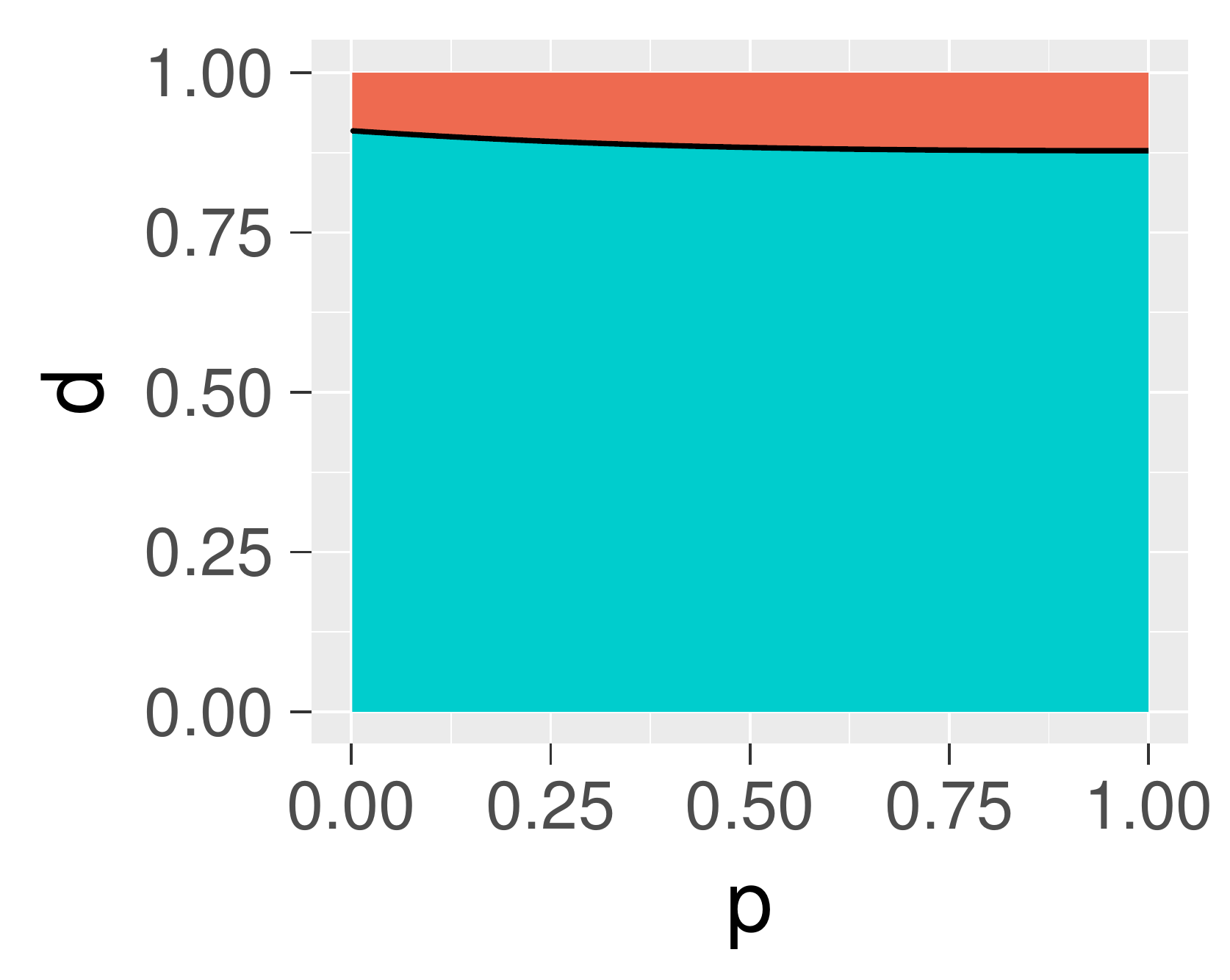}
 \caption{{\bf Phase diagram of clique splitting with generalized modularity density at large external influence as $\chi$ is varied}. The values of clique size ratio $p$ and link density $d$ where the M phase occurs is shown in orange and where the S phase occurs is shown in blue. Results are for different values of the control parameter $\chi$: (a) $\chi=0$, (b) $\chi=1$, (c) $\chi=3$, (d) $\chi=10$. The external influence parameter is $t=10^6$}
 \label{fig:Qgchi}
\end{figure}

\begin{figure}
\centering
(a)\includegraphics[width=0.45\textwidth]{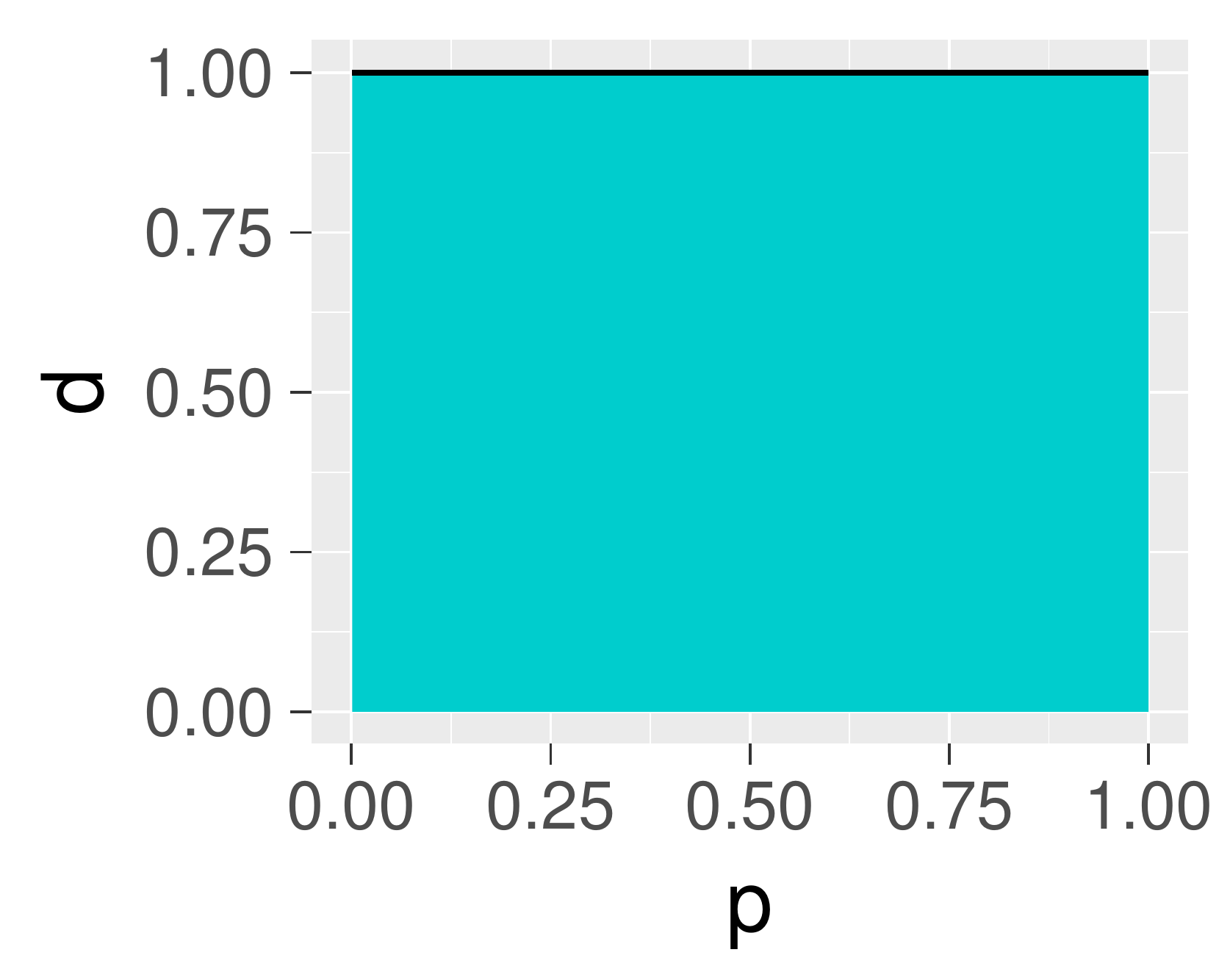}
(b)\includegraphics[width=0.45\textwidth]{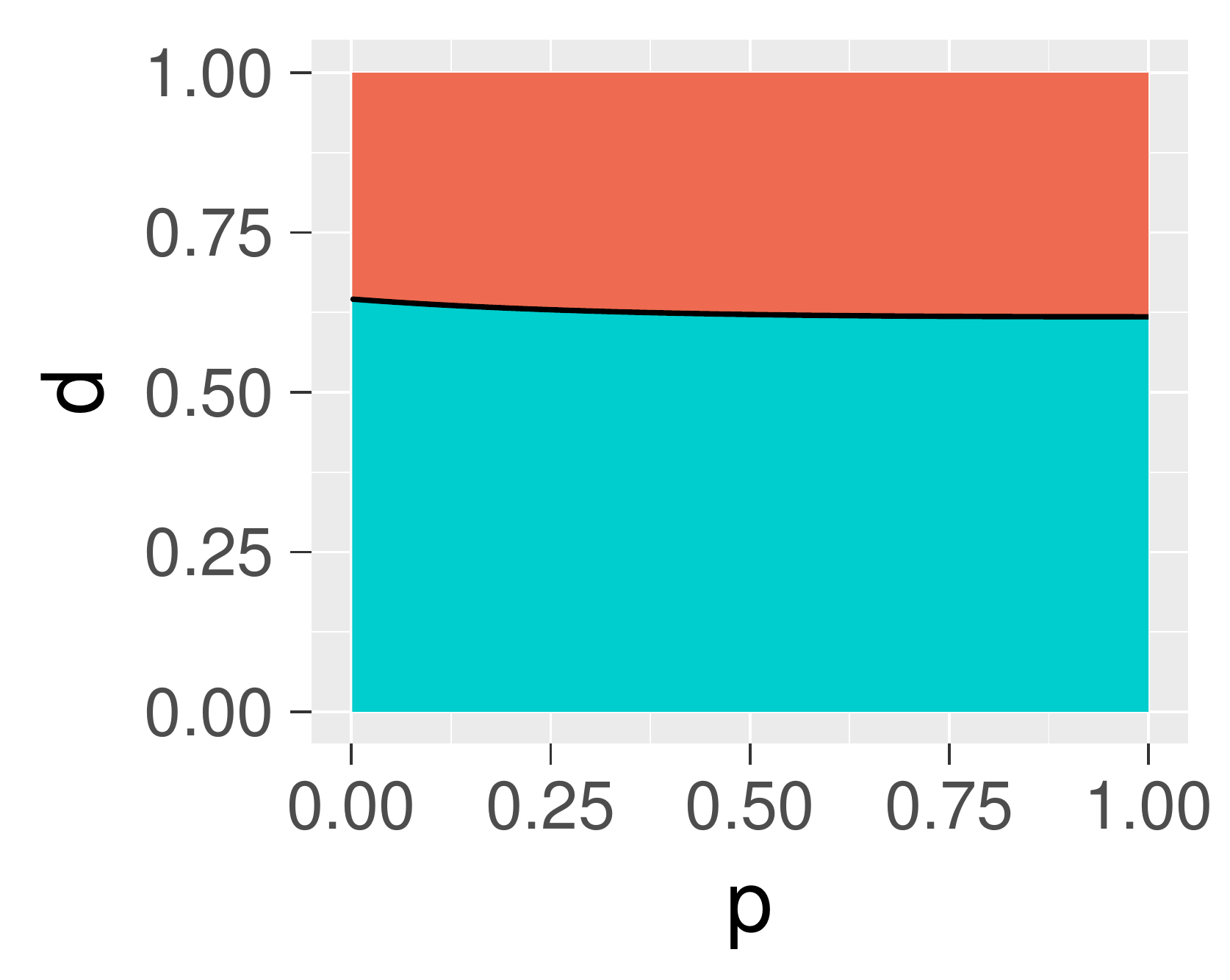}
(c)\includegraphics[width=0.45\textwidth]{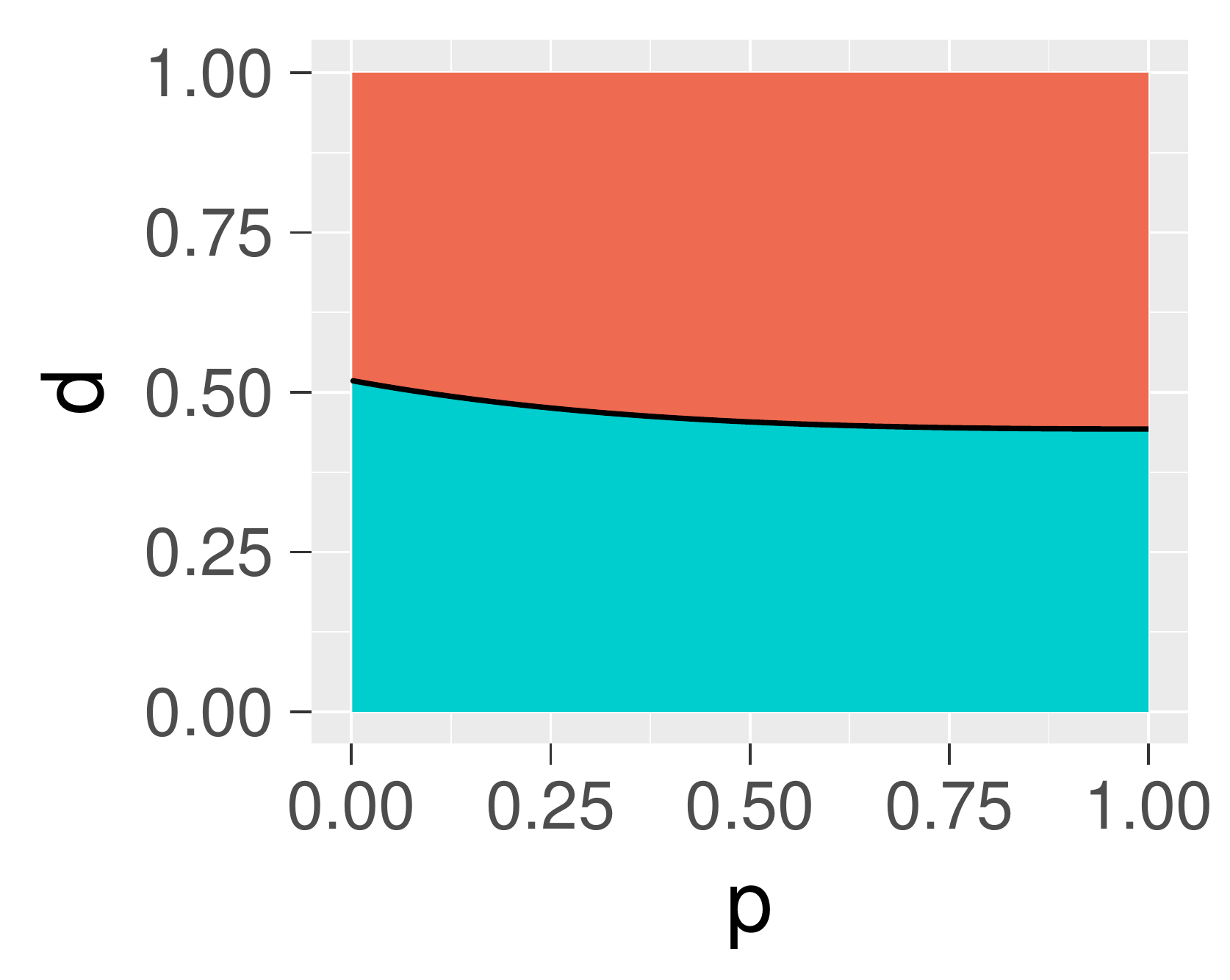}
(d)\includegraphics[width=0.45\textwidth]{Qg10-6.pdf}
\caption{{\bf Phase diagram of clique splitting with generalized modularity density at fixed $\chi$ as the external influence is varied}. The values of
 clique size ratio $p$ and link density $d$ where the M phase occurs is shown in orange and where the S phase occurs is shown in blue. Results are for $\chi = 1$ and different choices of the external influence parameter $t$: (a) $t=0$, (b) $t=1$, (c) $t=10$, (d) $t=10^6$.}
 \label{fig:Qgphase}
\end{figure}

In the limit of large external influence parameter $t$, which is often the situation encountered in empirical studies where RL problems are considered problematic, 
the value of $\delta_{Q_g}$ 
for a given value of $\chi$ is
\begin{equation}
    \label{equ:Qgworst}
    \lim_{t\rightarrow\infty}\delta_{Q_g}=\frac{r}{2}\left[\left(1+\frac{2}{r}\right)^{\frac{\chi}{\chi+1}}-1\right]\; .
\end{equation}
This limit increases from $\delta_{Q_g}=0$, when $\chi=0$, to $\delta_{Q_g}=1$, when $\chi\rightarrow \infty$, for all values of $p$.
At intermediate values of $\chi$ the result is only weakly dependent on $p$, being just slightly larger at small $p$, as can be seen in Fig.~\ref{fig:Qgchi}.
The figure shows shows the phase diagram as a function of $p$ and $d$ at various values $\chi$ for large $t$.
For $\chi = 0$, when $Q_g=Q$, the cliques are merged at all values of $p$ and $d$ as shown in Fig.~\ref{fig:Qgchi}(a). 
For $\chi > 0$ at smaller values of $d$ the cliques separate and are, thus, resolved.
As $\chi$ increases, $\delta_{Q_g}$ also increases and approaches 1 in the limit of large $\chi$, Figs.~\ref{fig:Qgchi}(b)-(d), meaning that at large $\chi$ the cliques are always resolved. 

The effect of varying $t$ at fixed $\chi$ on the ($p$, $d$) phase diagram are shown in Fig.~\ref{fig:Qgphase}. As shown in Fig.~\ref{fig:Qgphase}(a), at $t=0$ when there is no influence by the external component on the two cliques, the S phase occupies the entire space and $\delta_{Q_g}=1$ for all $p$.
In this case, the cliques are always separated unless they are fully connected to each other.
For $t>0$, when there is some influence from an external component, the cliques are merged and, thus, not resolved for large values of $d$. As $t$ increases, shown in Figs.~\ref{fig:Qgphase}(b)-(d), the M phase occupies an increasing area and $\delta_{Q_g}$ decreases until reaching the limiting value given by Eq.~\ref{equ:Qgworst}.

\begin{figure}
    \centering
    \includegraphics[width=0.7\textwidth]{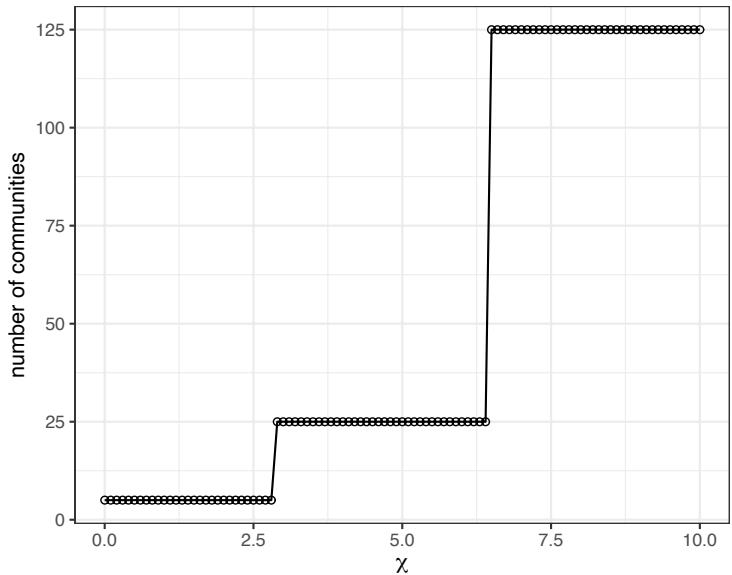}
    \caption{{\bf Number of communities found using $Q_g$ with different $\chi$.} Communities in each level (from the largest to the smallest) in the hierarchy are revealed as $\chi$ is varied.}
    \label{fig:levelplot}
\end{figure}

These results show that, as the control exponent $\chi$ is varied, a wide range of $\delta_{Q_g}$ results. The range increases 
with $t$ and varies from 0 to 1, the complete possible range, in the limit of large $t$.
This freedom gives leeway in applications to choose $\chi$ so that $\delta_{Q_g}$ matches the expected critical resolution link density $\delta_{\rm exp}$.

In general, as $\chi$ increases the number of communities found also increases, but gives stable results for a range of $\chi$. (See the example discussed in Sec.~\ref{sec:artificial network}.)
Increasing $\chi$ thus tends to result in smaller communities being detected.
The appropriate, or best, choice of $\chi$ depends on the problem. 
If there is some ``ground truth'' knowledge about the community structure in the network, or in similar networks, that knowledge can be used to select a $\chi$ that results in communities that match the ground truth.
If there is no ground truth knowledge, then a default choice of $\chi=1$ may be appropriate. That choice results in a critical resolution density of $\delta_{Q_g}=1/2$ in the limit of large $t$ and $r$ (Eq.~\ref{equ:Qgworst}).
Thus, an advantage of $Q_{g}$ is that even for the extreme values of $t$ and $r$, the metric has a positive lower bound of $\delta_{Q_g}$ that can be controlled by $\chi$.

In contrast to $Q$~\cite{newman2002}, $Q_{ds}$~\cite{mingming2013}, $Q_{x}$~\cite{chen2018} and $Q_{AFG}$~\cite{arenas2008} (see Supplemental Information~\ref{sub:metric}), $Q_g$ has a finite non-zero lower limit of $\delta$, which implies that for $d$ smaller than this value, the two cliques of the benchmark network are guaranteed to be split for all possible values of  $(r,t)$. Thus, $Q_g$ can successfully avoid resolution limit problem in these extreme cases (See last paragraph in Section~\ref{sec:resolutionscale}).
While the metric $Q_w$ (see Supplemental Information~\ref{sub:metric}) also shows this lower limit (Table~\ref{tab:table1}), the advantage of $Q_g$ is that the lower limit of $\delta_{Q_g}$ can be adjusted by tuning the parameter $\chi$ for any desired resolution density. 
Table~\ref{tab:table1} summarizes the kind of resolution problems with $Q$, $Q_{ds}$, $Q_x$, $Q_w$ and $Q_{AFG}$ that would be encountered when tested on the benchmark network (See Supplemental Information Section~\ref{sub:diagram} for details).

In principle, a reasonable $\delta_{\rm exp}$ is always in $[0,1]$ but a given metric can still have a $\delta$ that is out of this range. Since $\delta_{\rm exp}$ is strictly positive (no matter how small), if it is possible to construct a network for which $\delta\rightarrow0$ then that metric presents a resolution limit problem. Even worse, if $\delta < 0$, it would result in merging of disconnected communities. On the other hand $\delta\rightarrow1$ does not pose a resolution problem as long as $\delta\le1$ and $\delta\ge\delta_{\rm exp}$ is satisfied. However, higher $\delta_{\rm exp}$ imposes a stricter criterion for merging. But if $\delta>1$, it will have the unwanted consequence of cliques being subdivided. Thus, a metric is problematic if it can not avoid $\delta\rightarrow0$, $\delta<0$ or $\delta>1$.

\begin{table}[]
    \centering
    \begin{tabular}{||c|c||}
    \hline
        {\bf Metric} & {\bf Resolution limit problem} \\ [0.5ex] 
 \hline \hline
        $Q$ & $\delta\rightarrow0$ when $t\rightarrow\infty$\\
    \hline
    $Q_{ds}$ & $\delta<0$ when $p$ is small\\
    \hline
    $Q_x$ & $\delta<0$ when $p$ and $\rho$ are small  \\
    \hline
    $Q_w$ & $\delta_{min}=0.236$ when $t\rightarrow\infty$ and $p=1$ \\
    \hline
    $Q_{AFG}$ &$\delta<0$ when $s<0$ and $p$ is small  \\
    & $\delta>1$ when $s>0$ and $p$ is small\\
    \hline
    \end{tabular}
    \caption{{\bf Resolution limit problems of different metrics.} $Q,Q_{ds}$, $Q_x$, $Q_w$ and $Q_{AFG}$ have different resolution limits problems. $\rho$, which appears in $Q_x$, is the global link density. $s$ is used in the metric $Q_{AFG}$ as a weight to every node (equivalent to adding a self-loop to every node) and thereby modifying the strength of a community. $Q_{AFG}$ reduces to modularity at $s=0$, and by controlling $s$ substructures ($s>0$) or superstructures ($s<0$) can be explored.}
    \label{tab:table1}
\end{table}

\subsection{Applications}

\subsubsection{American college football network}

\begin{figure}
    \centering
    \includegraphics[width=1.0\textwidth]{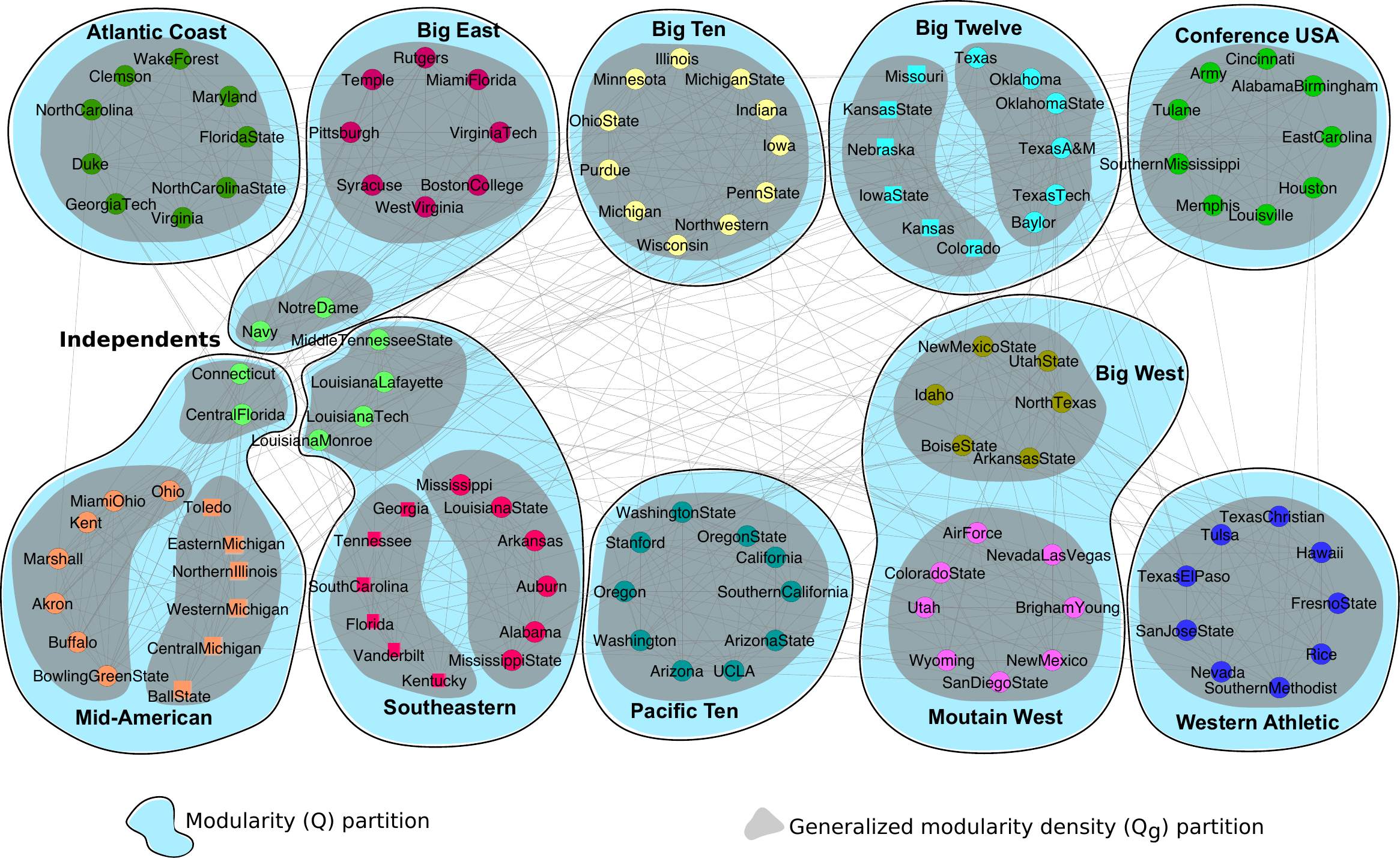}
    \caption{{\bf Communities found in American college football network.} Blue blobs show the communities detected by modularity and gray blobs show the communities found by generalized modularity density.}
    \label{fig:my_label}
\end{figure}

We use the $Q_g$ metric to detect communities in the network of American college football games between Division IA colleges during regular season of Fall 2000~\cite{newman2002, evans2010}. A link between two colleges is present if they played a game against each other. Colleges play games within the same conference more frequently, thus, a community detection algorithm should be able to recover these conferences from the network data. First, we show the result of using modularity ($Q$) that are indicated by light blue blobs in Fig~\ref{fig:my_label}. It matches the conference memberships (distinguished by node color) well except Independents, which are absorbed by three communities and that it groups Big West and Mountain West in the same community. Using $Q_g(\chi=3)$ in this network we find communities that are shown by gray blobs. There are some key differences between the $Q$ and $Q_g$ partitions. First, the $Q_g$ partition does not merge the Independents with other conferences. Instead, it divides them into three disjoint communities. Second, it successfully identifies the Big West and Mountain West as two different groups. But more interestingly, unlike modularity, it divides each of the Mid-American, Southeastern, and Big Twelve conferences into two communities. This apparent deviation from ground truth actually turns out to be a major advantage of using $Q_g$. Each of these three conferences have subdivisions within them that are in perfect agreement (considering their membership as of year 2000) with the partition found by $Q_g$. Mid-American conference has East Division and West Division, Southeastern also has Eastern Division and Western Division, whereas Big-Twelve conference has Northern Division and Southern Division. These subdivisions are indicated by different node shapes (circles and squares) in Fig.~\ref{fig:my_label}.

\subsubsection{Artificial network with hierarchical community structure}
\label{sec:artificial network}
\begin{figure}
    \centering
    \includegraphics[width=1.0\textwidth]{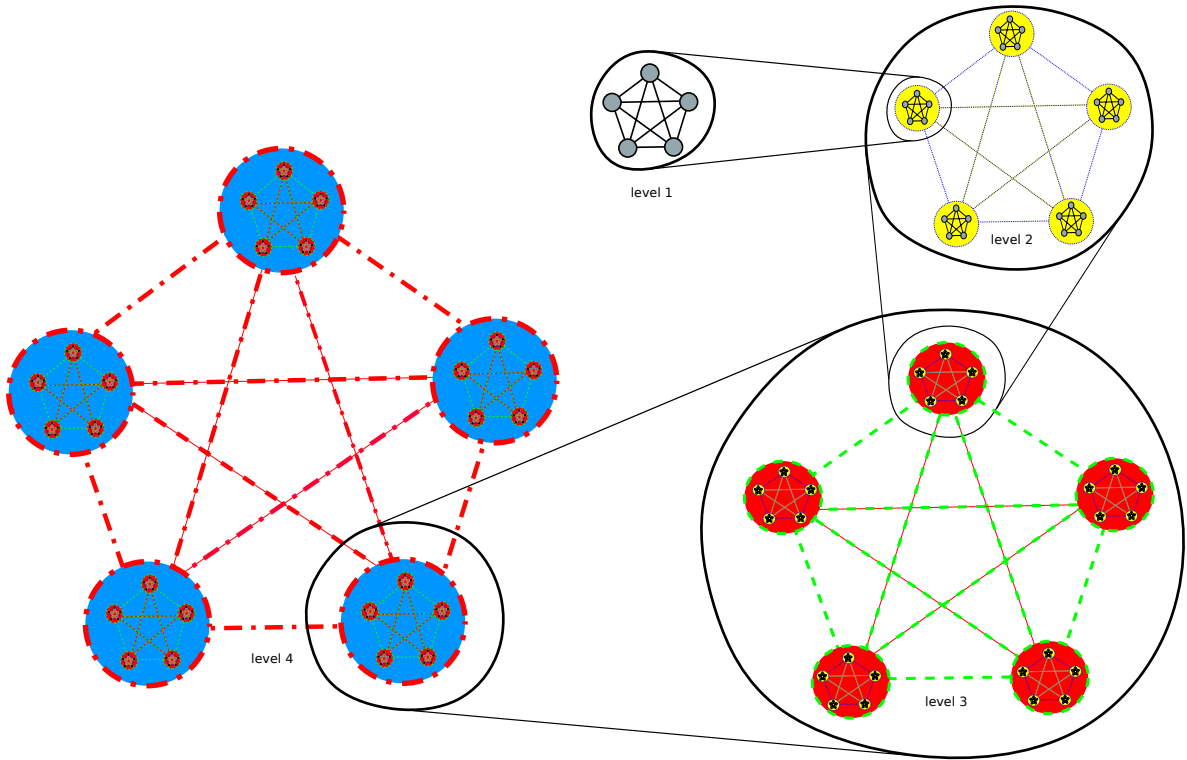}
    \caption{{\bf Example hierarchical network.} Level 1: A clique of five nodes. Level 2: A clique of five level 1 cliques. Level 3: A clique of five level 2 cliques. Level 4: A clique of five level 3 cliques.}
    \label{fig:multilevel}
\end{figure}

To demonstrate the ability of $Q_g$ for detecting the community structure at different resolution densities, we construct a hierarchical network. Similar constructions have been used as a model for hierarchical network structure~\cite{arenas2008}. We consider a structure shown in Fig.~\ref{fig:multilevel} that includes four levels of hierarchy, although it can be extended to include any number of levels. The elementary level (level 1) is a clique formed by fully connecting five nodes with links weighted $\alpha_1$. To construct a level 2 network, we use the clique network from level 1 as a {\it generalized node} to form a clique of size 5 with links weighted $\alpha_2$. Link between two generalized nodes is achieved by connecting all the internal nodes from one generalized node to those in another generalized node. Similarly, level $k$ network is constructed by using level $k-1$ network as a generalized node to form a clique of size 5 with links weighted $\alpha_k$. Here we keep $\alpha_1>\alpha_2>...>\alpha_k$ so that the hierarchy of structure is preserved.\\
We use the metric $Q_g$ on a level 4 network with $\alpha_k=5-k$ and show that it successfully detects the planted hierarchical communities at every level. The level 4 network consists of 125 level 1 cliques, and 625 nodes in total. The results obtained by maximizing $Q_g$ is shown in Fig.~\ref{fig:levelplot}. We observe that the 5 level 3 cliques are detected when $\chi<2.8$, the 25 level 2 cliques are detected when $2.9<\chi<6.4$, the 125 level 1 cliques are detected when $\chi>6.5$. There are 3 stages, corresponding to 3 levels of construction. There should not be a single ``best" choice of $\chi$ by the nature of the problem. The choice of $\chi$ or desired resolution density should be based on specific requirement and the background information of the particular problem.

\section{Conclusion}

Community detection in networks is commonly performed by finding the partition of the network nodes that maximizes an objective function. Such a partition can sometimes yield unexpected community structure. Resolution limit, for example, is an unwanted but inevitable consequence of modularity maximization. Other such metrics, namely modularity density measures, which attempt to fix this problem also differ in the community structure that they obtain and can also violate our general expectation. While at what number of cross links between two strongly connected groups of nodes should be called a single community remains mostly subjective and vague, our metric $Q_g$ provides a quantifiable notion and solves the resolution limit problem. In particular, with a free parameter $\chi$, one can control this threshold of merging two cliques. It is quite appealing to have a metric that can be adjusted to meet the specific requirement set by the user because the idea of a community may vary from one application to another and may be specific to the network under consideration. At the same time, due to its ability to detect communities at many resolution densities it also useful in uncovering the hierarchical community structure, an inherent characteristic observed in many complex networks.
The existing benchmarks, e.g. the ring of cliques, are too restrictive to evaluate and compare the performance of different metrics with respect to solving the specific resolution limit problem. In this paper we consider a more general yet simple network structure, which can be used to quantitatively examine the limits of metrics such as modularity. Using this general framework we demonstrated that our metric $Q_g$ eliminates resolution limit problem at a desired resolution density, shows better performance, and is straightforward to extend for studying weighted and directed networks. Among other important problems, finding communities at high resolution is particularly useful in analyzing gene regulatory networks where the goal of functional annotation of genes is to find very specific gene functions~\cite{mentzen2008}. 

\section*{Acknowledgments}
This work was supported by the NSF through grants DMR-1507371 and IOS-1546858.

\pagebreak

\beginsupplement
\begin{center}
\title{Supplemental Information}
\end{center}

\section{Other metrics}
\label{sub:metric}
Besides the metric $Q_g$, we test the performance of the following metrics. Each variable has the same meaning as Eq.~\ref{eq:Qg} (in the main text) unless otherwise noted.\newline
Modularity~\cite{newman2002}:
\begin{equation}    \label{eq:Q}
Q=\frac{1}{2m}\sum_c \left(2m_c-\frac{K_c^2}{2m}\right)
\end{equation}
Weighted Modularity~\cite{haq2019}:
\begin{equation}     \label{eq:Qw}
Q_w=\frac{1}{2m}\sum_c \left(2m_c-\frac{K_c^2}{2m}\right)(\rho_c+1)
\end{equation}
Excess Modularity Density~\cite{chen2018}:
\begin{equation}     \label{eq:Qx}
Q_x=\frac{1}{2m}\sum_c \left[2m_c(\rho_c-\rho)-\frac{K_c^2(\rho_c-\rho)^2}{2m}\right]
\end{equation}
Here $\rho=2m/[n(n-1)]$ is the global link density.\\
Modularity Density introduced in Ref.~\cite{mingming2013} has a term that corresponds to Split Penalty. But as discussed in Ref.~\cite{chen2018}, this term may be problematic. Therefore, here we analyze a modified version of modularity density without the Split Penalty term:\\
\begin{equation}    \label{eq:Qds}
Q_{ds}=\frac{1}{2m}\sum_c   \left(2m_c\rho_c-\frac{K_c^2\rho_c^2}{2m}\right)
\end{equation}
AFG method of modularity $Q_{AFG}$ in Ref.~\cite{arenas2008} can have different resolution densities by assigning self-loop weighted $s$ to each node and tuning the value of $s$. It still finds the partition by maximizing modularity after assigning the self-loops.
\begin{equation}    \label{eq:Qafg}
Q_{AFG}=\frac{1}{2m+2Ns}\sum_c   \left[(2m_c+2n_cs)-\frac{(K_c+2n_cs)^2}{2m+2Ns}\right]
\end{equation}
where $N$ is total number of nodes, $n_c$ is the number of nodes of community c.

\section{Derivation of equation of phase for modularity}
\label{sub:derive}
We assess the performance of modularity $Q$ using the benchmark test described in the main text.
In the form shown in Eq.~\ref{eq:Q}, modularity is the sum of the quantity within parenthesis over each community. Thus, two partitions of splitting or merging the two cliques yield the following values of modularity $Q$
\begin{equation}
    Q_{split}=Q_1+Q_2+Q_{ex}
\end{equation}
\begin{equation}
    Q_{merge}=Q_{(1+2)}+Q_{ex}
\end{equation}
where $Q_1, Q_2$ are the two terms corresponding to clique 1 and 2 as separate communities, $Q_{(1+2)}$ is the corresponding term when the two cliques are merged. $Q_{ex}$ is the sum over the remaining communities in the external component, which do not change in the two partitions. The difference between the two modularity values is given by
\begin{equation}
\Delta Q=Q_{merge}-Q_{split}=Q_{(1+2)}-Q_1-Q_2.
\end{equation}
Using Eq.~\ref{eq:Q}, we have:
\begin{equation}
    \label{eq:dQ0}
    \Delta Q=\frac{1}{2m}\left[(2m_{(1+2)}-\frac{K_{(1+2)}^2}{2m})-(2m_1-\frac{K_1^2}{2m}+2m_2-\frac{K_2^2}{2m})\right]
\end{equation}
According to the construction of the example network, we can rewrite Eq.~\ref{eq:dQ0} with $(n_1,n_2,m_{12},m_a,n_a,m_{1a},m_{2a})$. To simplify the expression and capture the principle features, we take $n_1, n_2 >> 1$. Recall the construction of separation of the two cliques from the external component, we also have $n_1^2>>m_{1a},n_2^2>>m_{2a}$. Plugging all in $\Delta Q$, we obtain:
\begin{equation}
    \label{eq:dQ}
    \Delta Q=\frac{1}{2m}\left(2m_{12}-\frac{2(n_1^2n_2^2+(n_1^2+n_2^2)m_{12}+m_{12}^2)}{n_1^2+n_2^2+2m_a+2m_{12}}\right)
\end{equation}

Eq.~\ref{eq:dQ} can be rewritten more concisely by omitting the normalization factors and using variables $(d,r,t)$ defined in Section~\ref{sec:benchmark}
\begin{equation}
    \label{eq:dQ1}
    \Delta Q\sim2d-\frac{2(1+rd+d^2)}{r+2d+2t}
\end{equation}
The space is reduced to three principal dimensions $(d,r,t)$, where $0\leq d \leq1$, $r\geq 2$ and $t\geq 0$. Eq.~\ref{eq:dQ1} is the equation of phase that is used to plot the phase diagram of Fig.~\ref{fig:Qphase}. 
We obtain $\delta_Q$, which determines the phase boundary as the value of $d$ for which $\Delta Q = 0$,
\begin{equation}
    \label{eq:Qphase}
    \delta_Q=\sqrt{t^2+1}-t
\end{equation}

\section{Benchmark test}
\label{sub:diagram}
By carrying out similar mathematical analyses as in the last section, we can obtain an equation of phase for each metric and we can identify possible RL problems of each metric. Some advantages of using this benchmark test includes that it covers a wider range of cases, it can be used by working on the formula without any guess or speculation of specific network, and it provides a clear view of metric performance including all RL problems previously reported. 
The difference between values of a metric between merge and split cases, for other metrics can be written as follows. \newline 
For weighted modularity $Q_w$ 
\begin{equation}
    \Delta Q_w= \frac{2r+2d+2}{r+2}(2d+r-\frac{(r+2d)^2}{r+2d+2t})-2(r-\frac{r^2-2+2d^2+2dr}{r+2d+2t}).
\end{equation}
For excess modularity density $Q_x$ 
\begin{equation}
\begin{aligned}
\Delta Q_x=  (2d+r)(\frac{2d+r}{r+2}-\rho)-\frac{(r+2d)^2}{r+2d+2t}(\frac{r+2d}{r+2}-\rho)^2 \\ - \left(r(1-\rho)-\frac{r^2-2+2d^2+2dr}{r+2d+2t}(1-\rho)^2\right).
\end{aligned}
\end{equation}
For modified (without the split penalty term) modularity density $Q_{ds}$ 
\begin{equation}
    \Delta Q_{ds}=\frac{(2d+r)^2}{r+2}-\frac{(2d+r)^4}{(r+2d+2t)(r+2)^2}-(r-\frac{r^2-2+2d^2+2dr}{r+2d+2t})
\end{equation}
For $Q_x$ (Eq.~\ref{eq:Qx}), in addition to $(d, r, t)$, the phase space consists of an extra principal variable $\rho$, which is the global link density and its maximum value $\rho_{max}$ is obtained when $n_a$ is smallest as other variables $(n_1, n_2, m_{12}, m_a)$ are fixed.

\begin{figure}
\centering
(a)\includegraphics[width=0.45\textwidth]{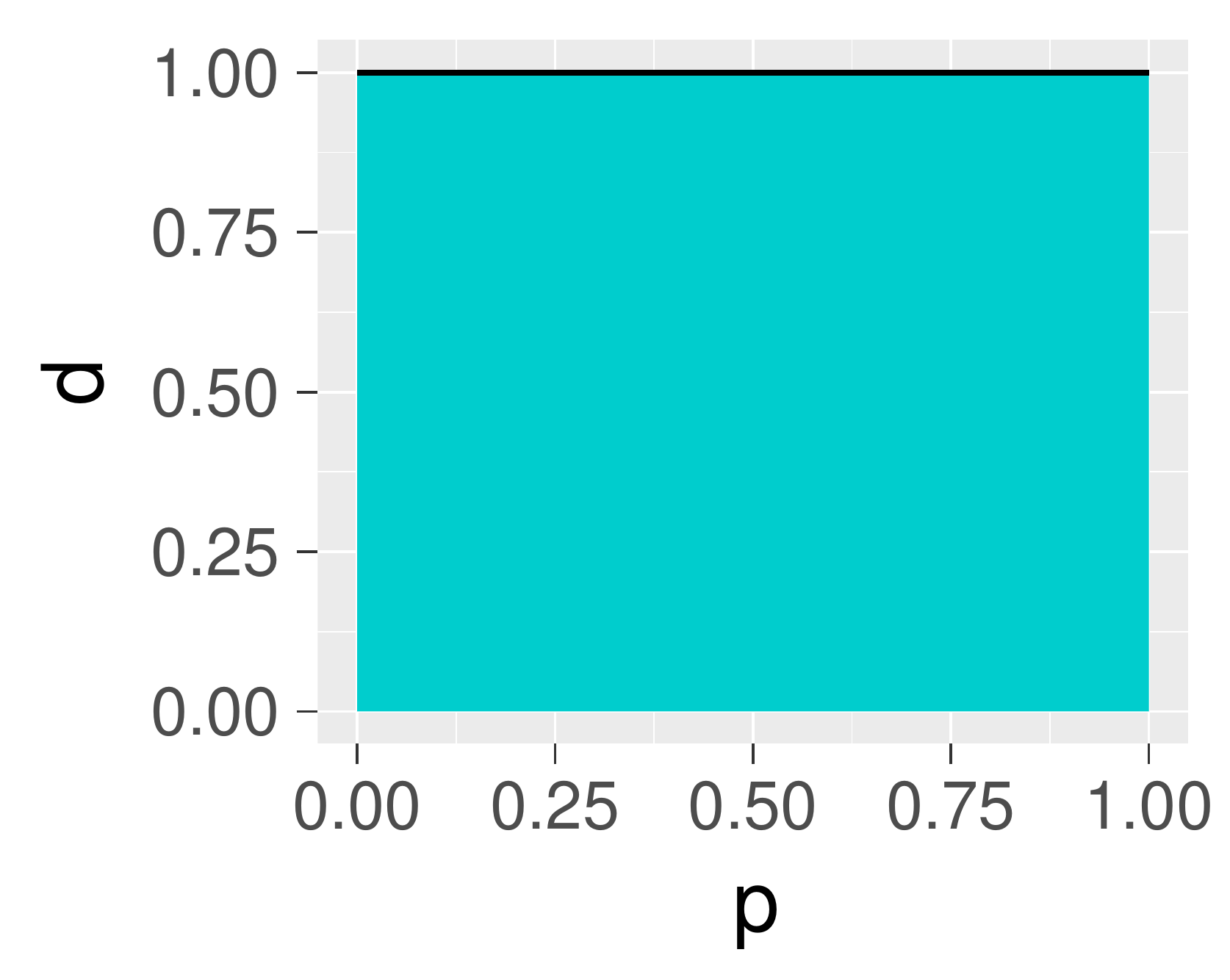}
(b)\includegraphics[width=0.45\textwidth]{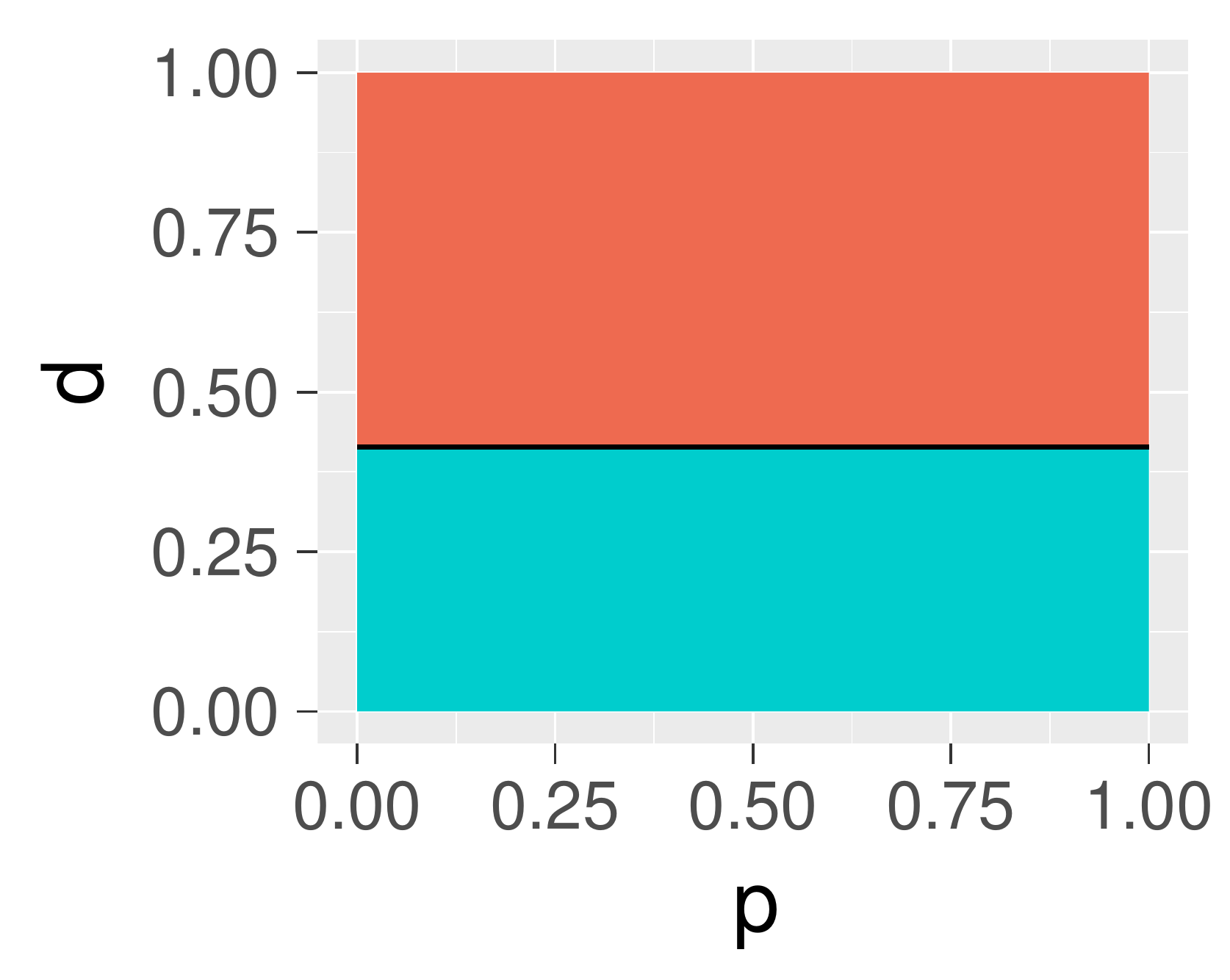}
(c)\includegraphics[width=0.45\textwidth]{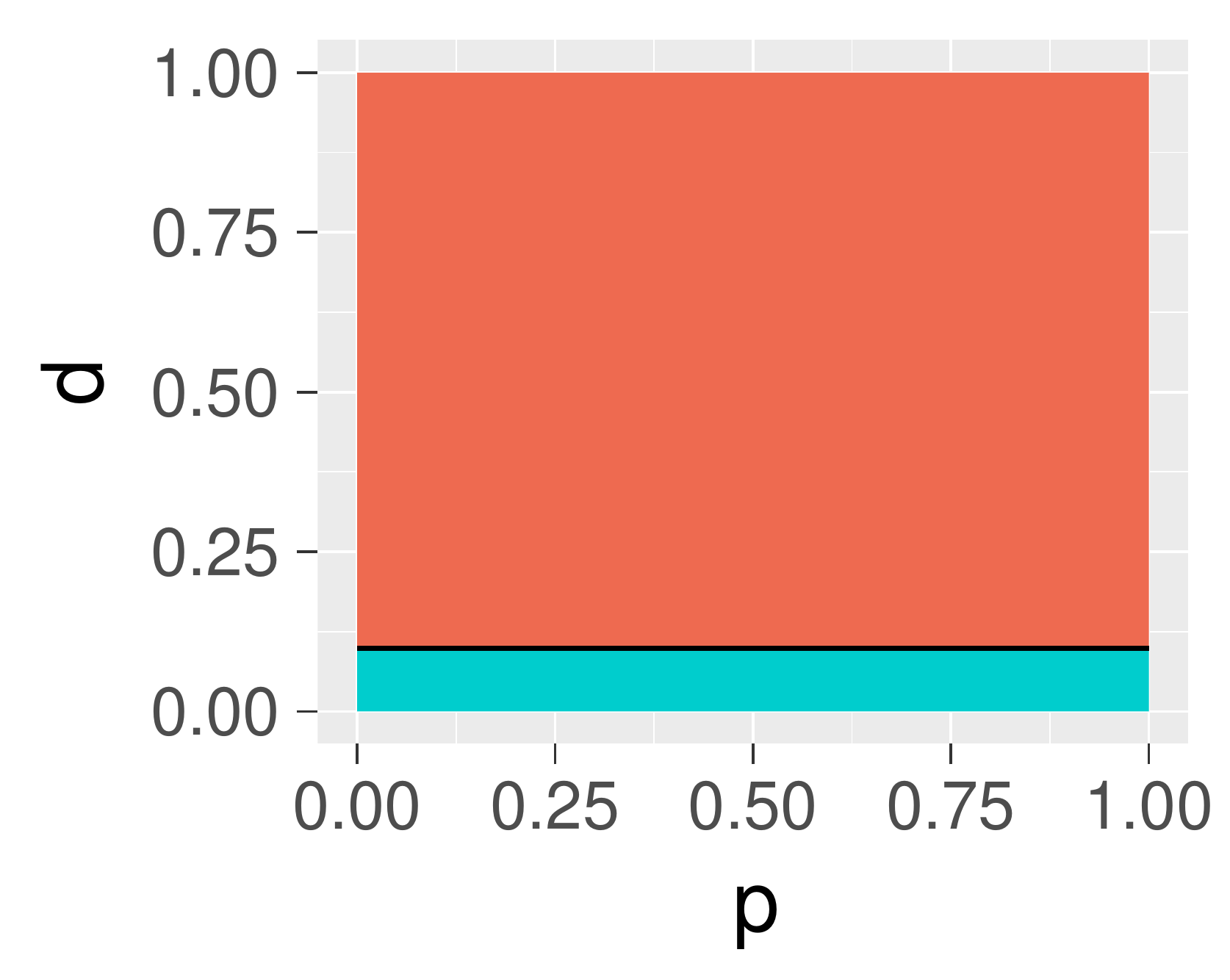}
(d)\includegraphics[width=0.45\textwidth]{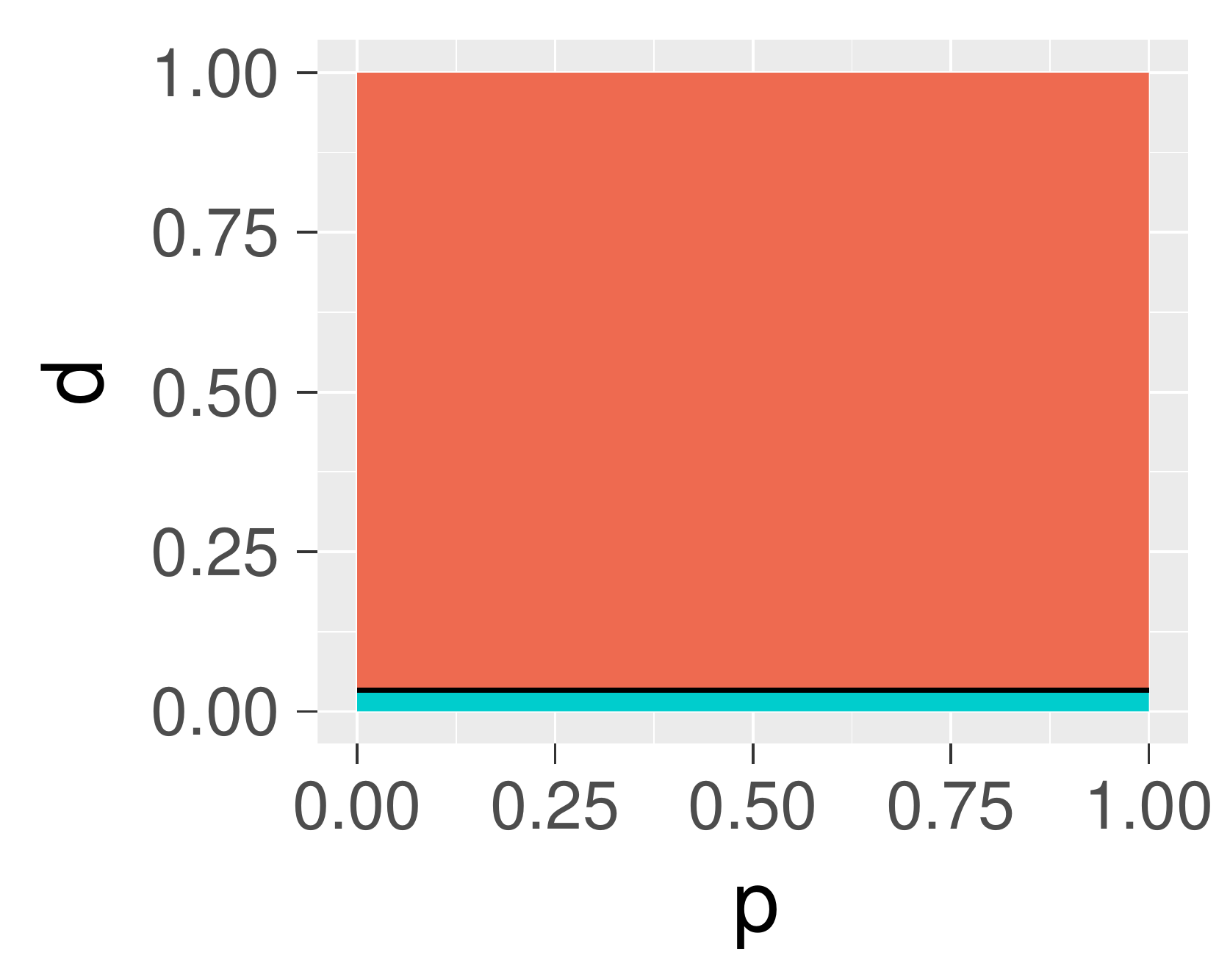}
 \caption{{\bf Phase diagram of clique splitting with modularity $Q$ as the external influence is varied.} The values of
 clique size ratio $p$ and link density $d$ where the M phase occurs is shown in orange and where the S phase occurs is shown in blue. Results are for different choices of the external influence parameter $t$: (a) t=0 (b) t=1 (c) t=5 (d) t=15.}
 \label{fig:Qphase}
\end{figure}

\begin{figure}
\centering
(a)\includegraphics[width=0.45\textwidth]{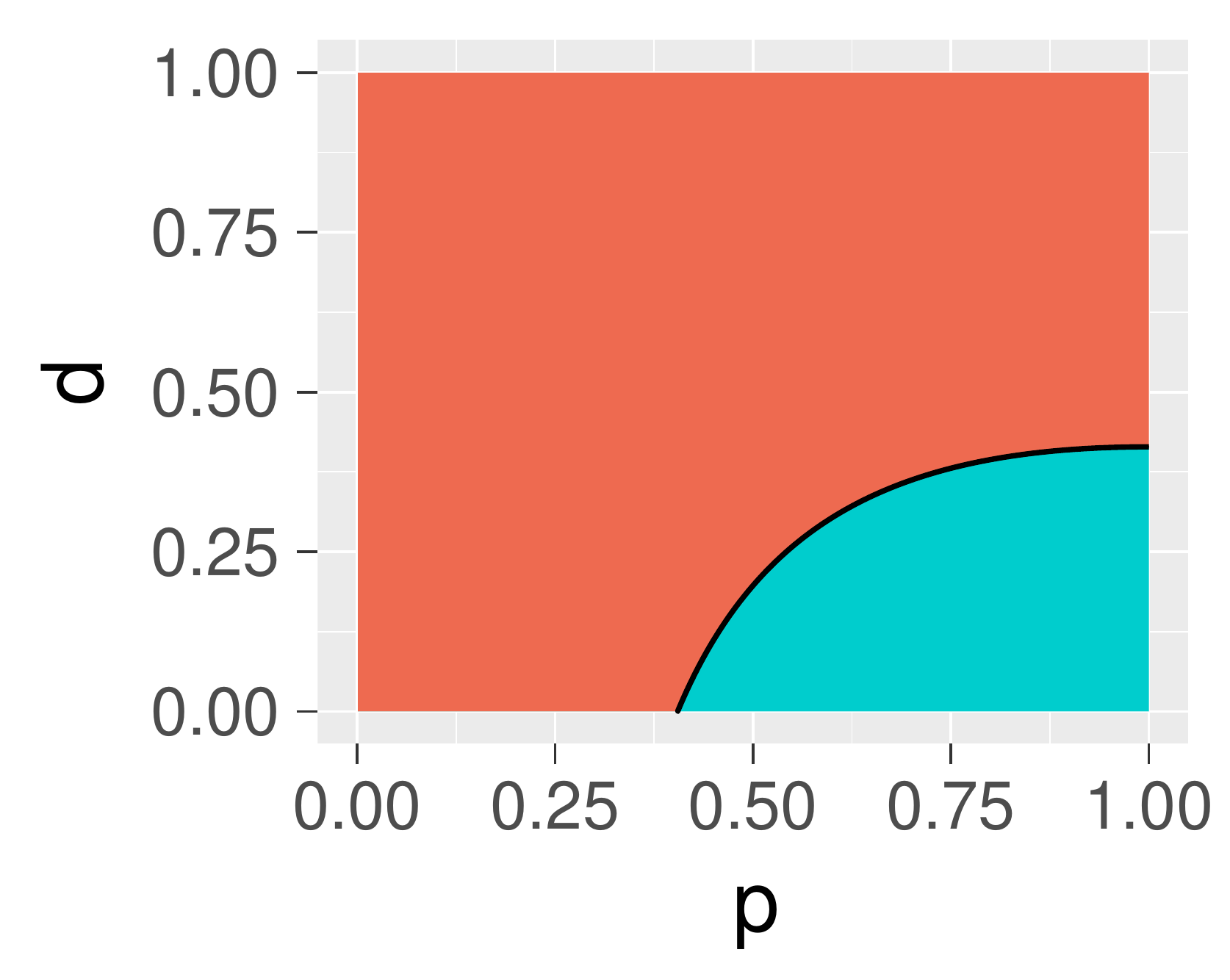}
(b)\includegraphics[width=0.45\textwidth]{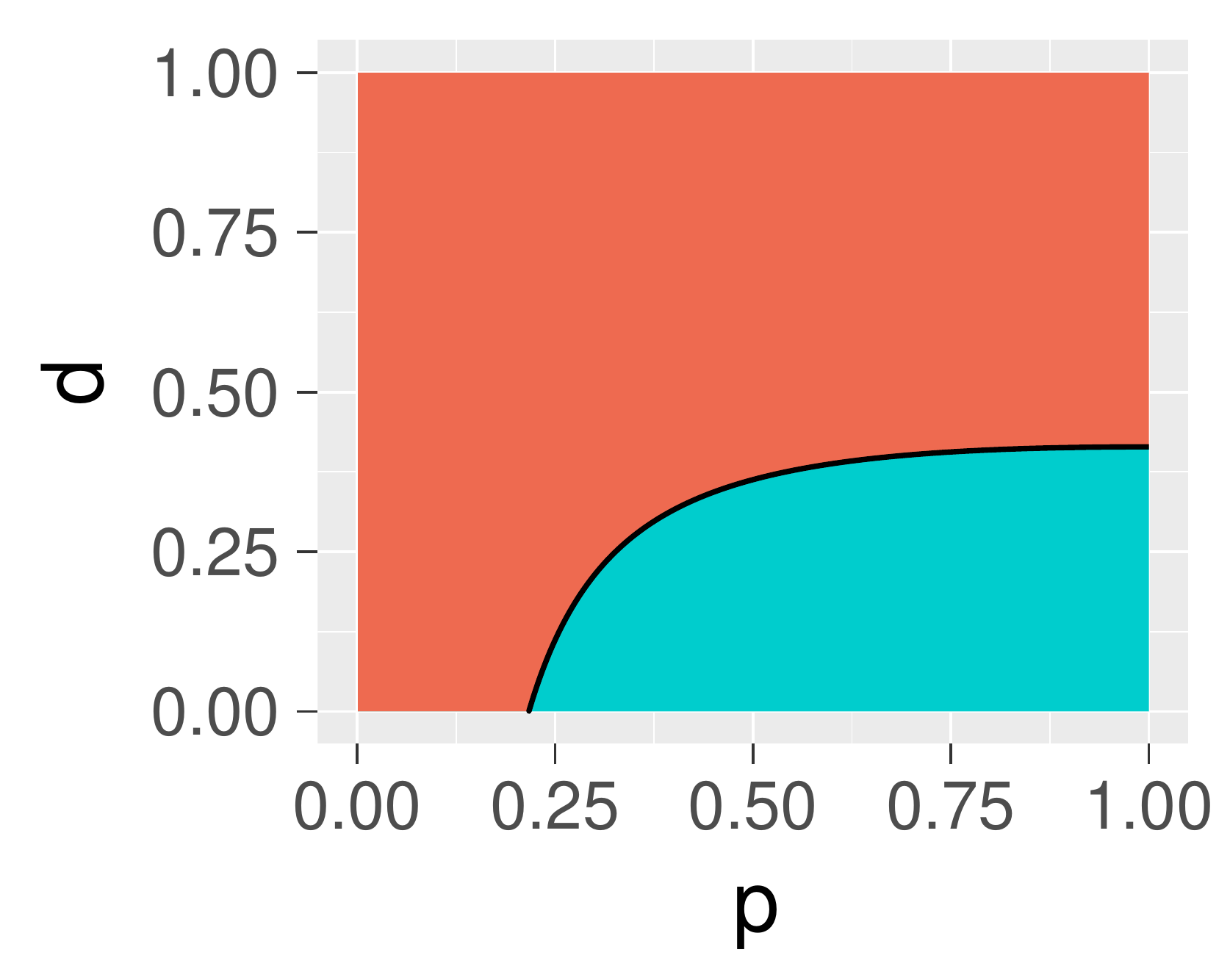}
(c)\includegraphics[width=0.45\textwidth]{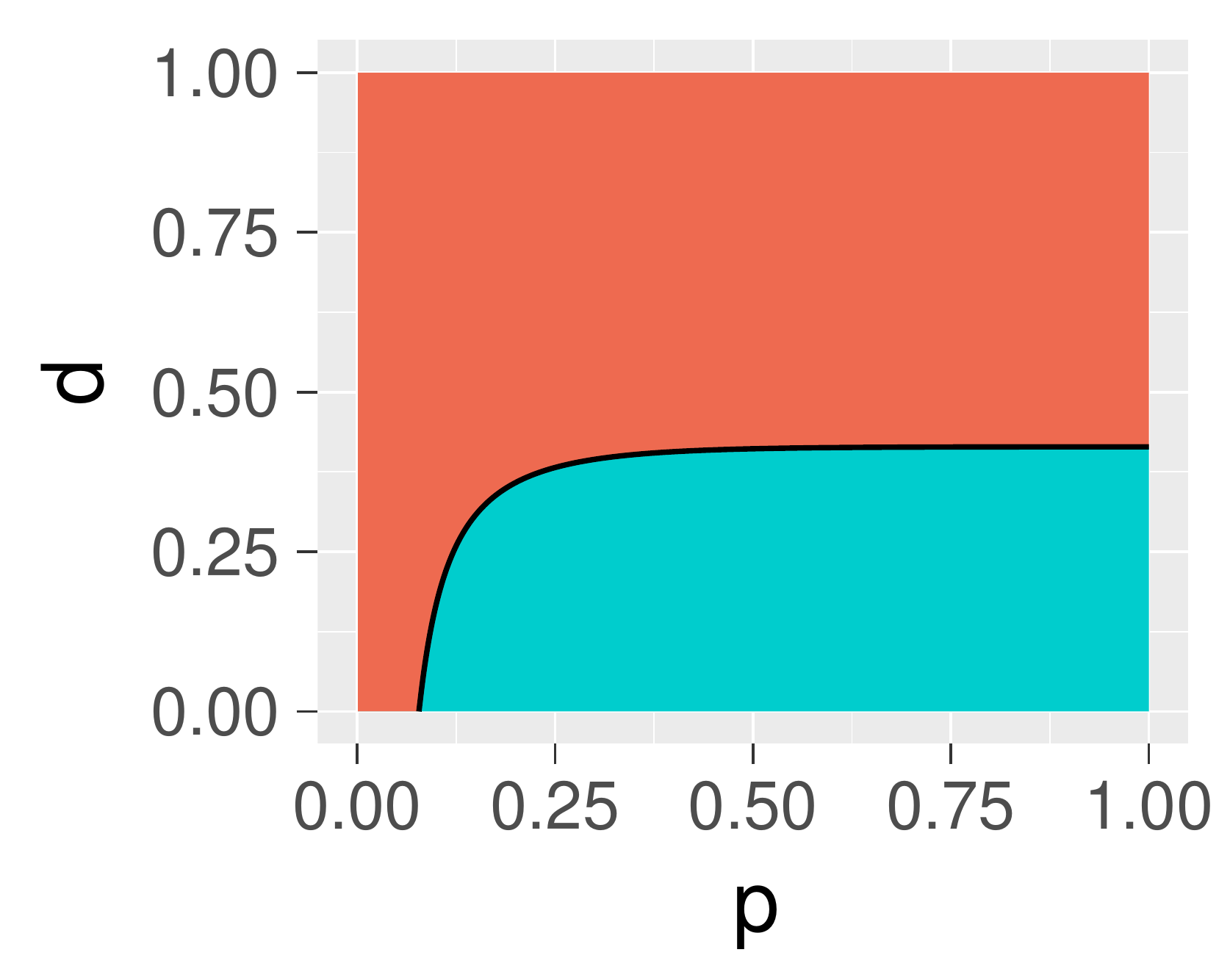}
(d)\includegraphics[width=0.45\textwidth]{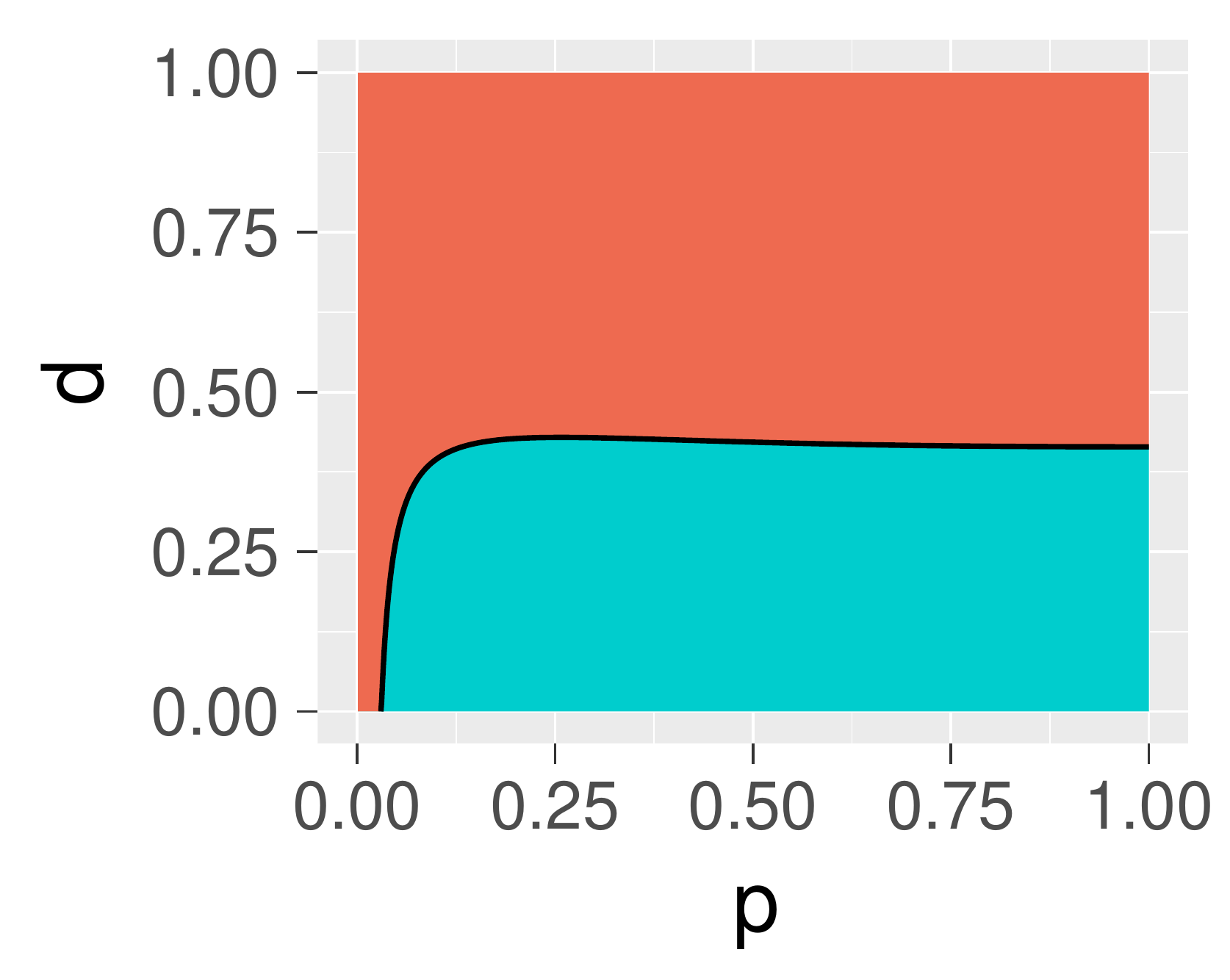}
 \caption{{\bf Phase diagram of clique splitting with modularity density $Q_{ds}$ as the external influence is varied.} The values of
 clique size ratio $p$ and link density $d$ where the M phase occurs is shown in orange and where the S phase occurs is shown in blue. Results are for different choices of the external influence parameter $t$: (a) t=0 (b) t=1 (c) t=5 (d) t=15.}
 \label{fig:Qdsphase}
\end{figure}

\begin{figure}
\centering
(a)\includegraphics[width=0.45\textwidth]{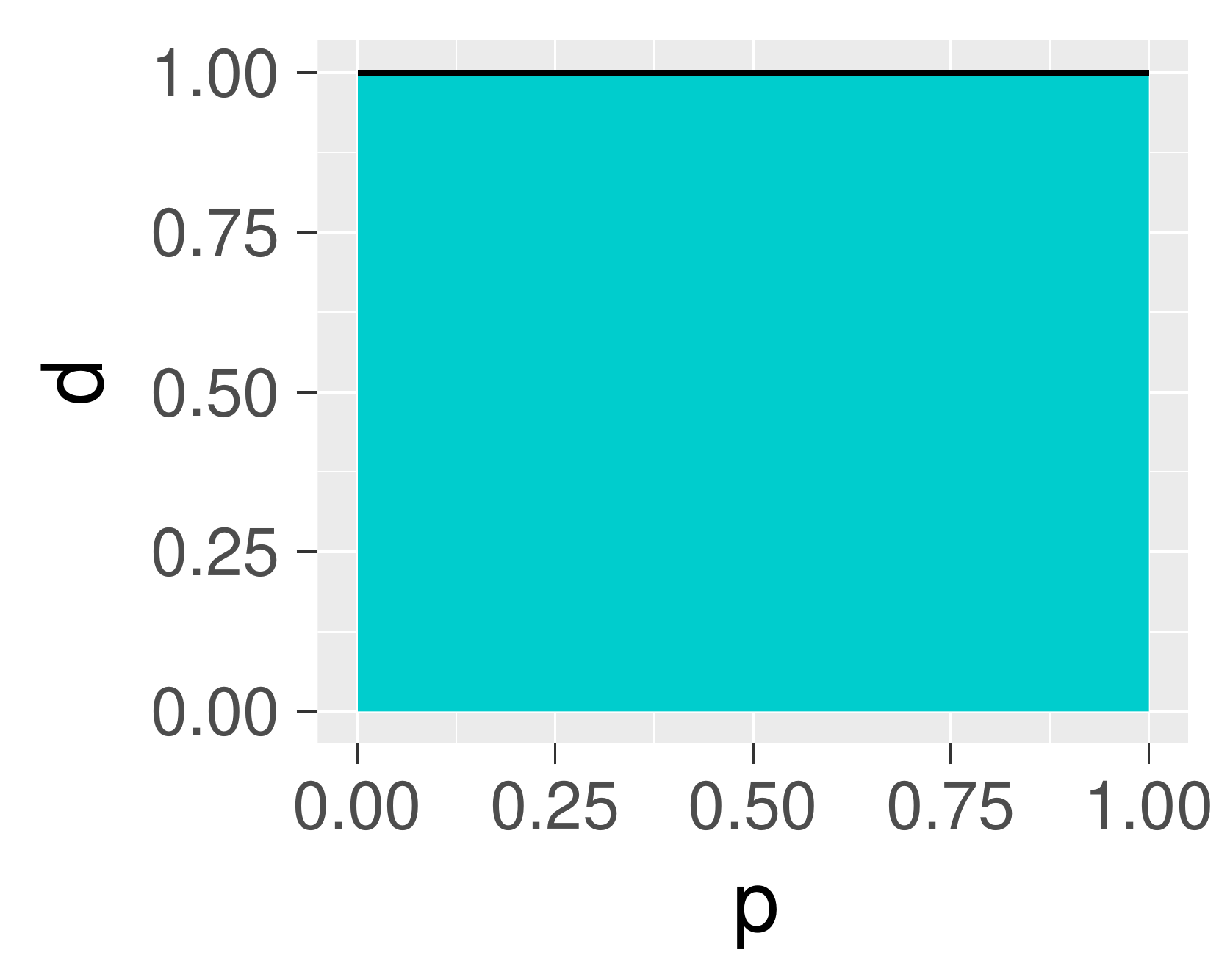}
(b)\includegraphics[width=0.45\textwidth]{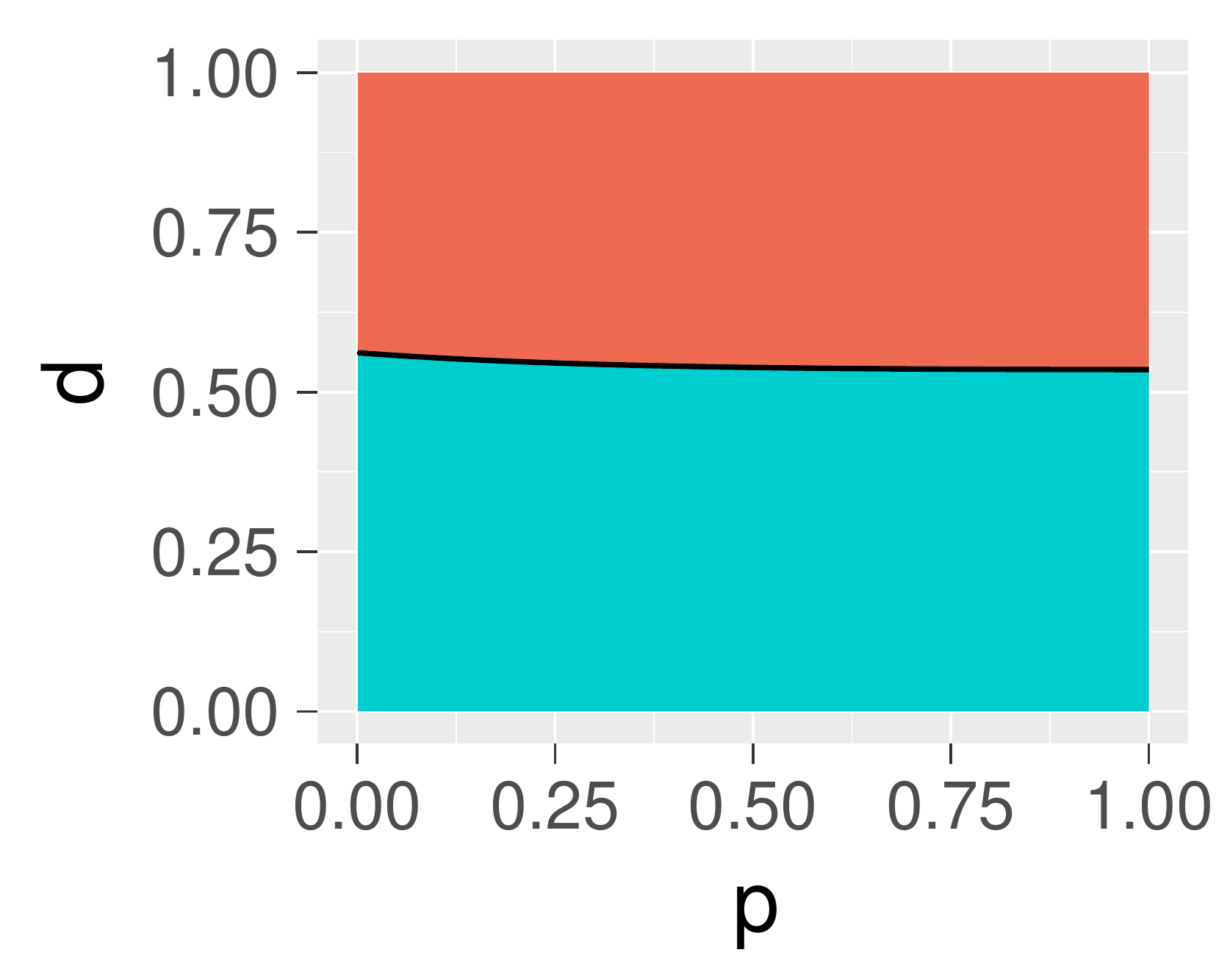}
(c)\includegraphics[width=0.45\textwidth]{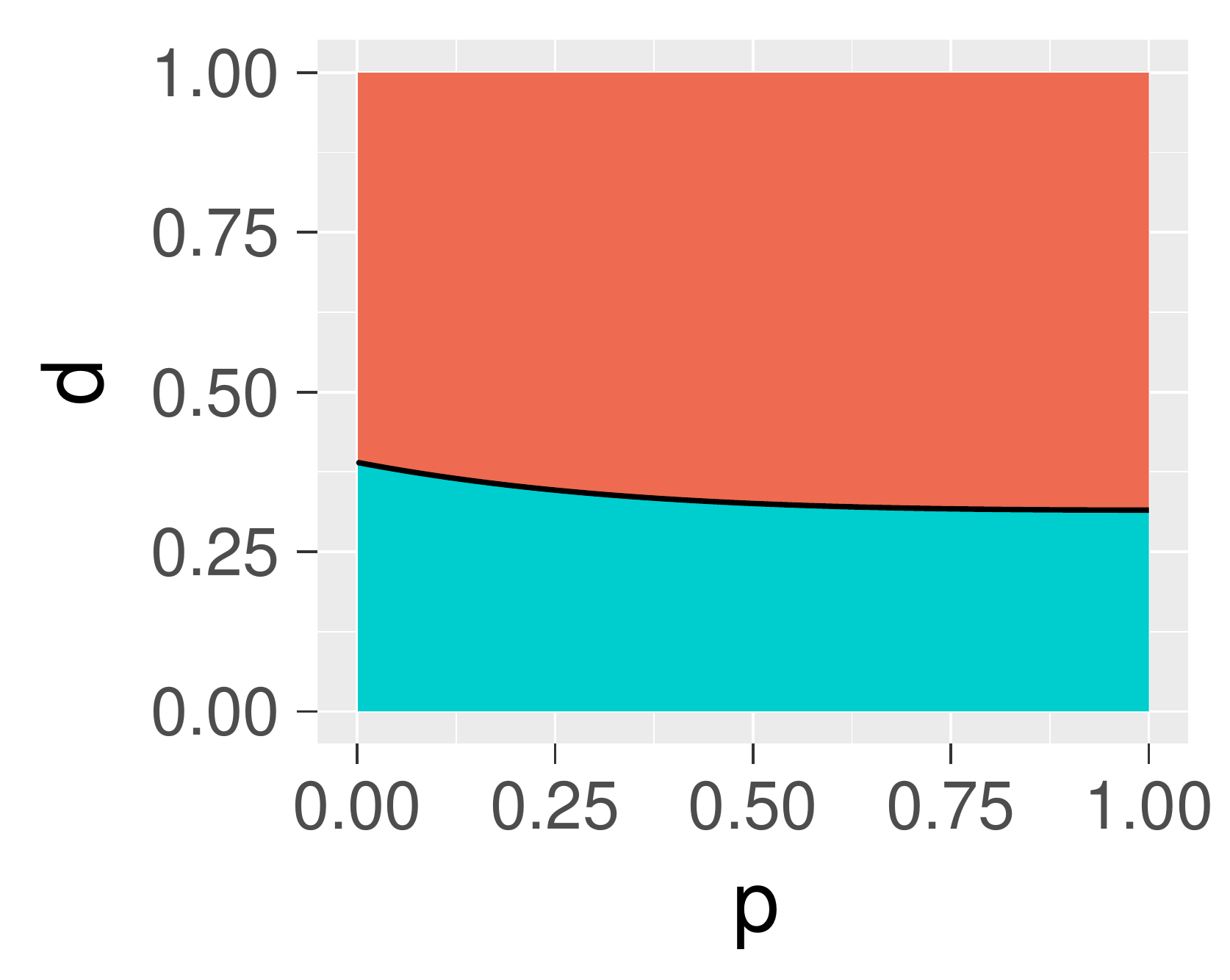}
(d)\includegraphics[width=0.45\textwidth]{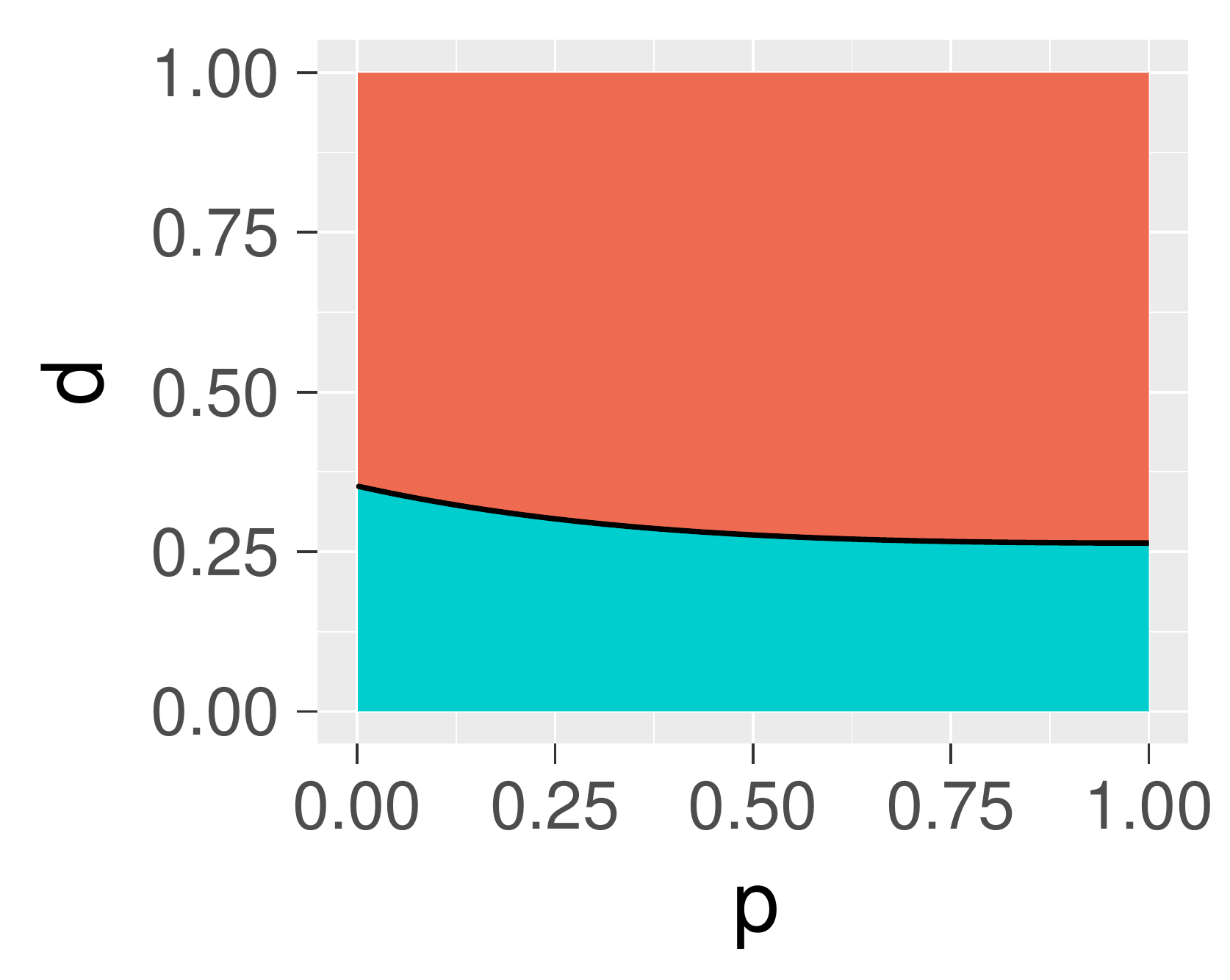}
 \caption{{\bf Phase diagram of clique splitting with weighted modularity $Q_w$ as the external influence is varied.} The values of
 clique size ratio $p$ and link density $d$ where the M phase occurs is shown in orange and where the S phase occurs is shown in blue. Results are for different choices of the external influence parameter $t$: (a) t=0 (b) t=1 (c) t=5 (d) t=15.}
 \label{fig:Qwphase}
\end{figure}

\begin{figure}
\centering
(a)\includegraphics[width=0.45\textwidth]{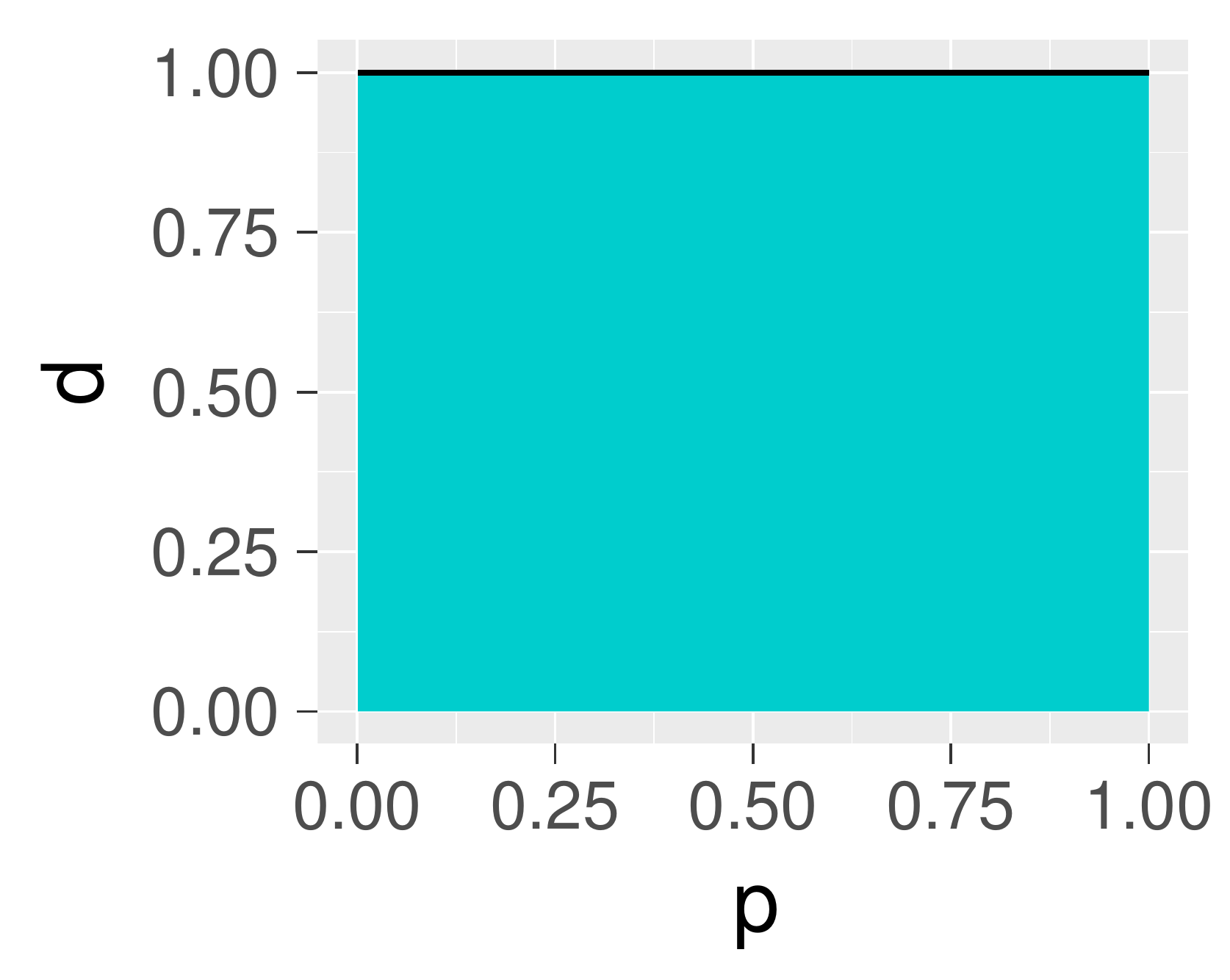}
(b)\includegraphics[width=0.45\textwidth]{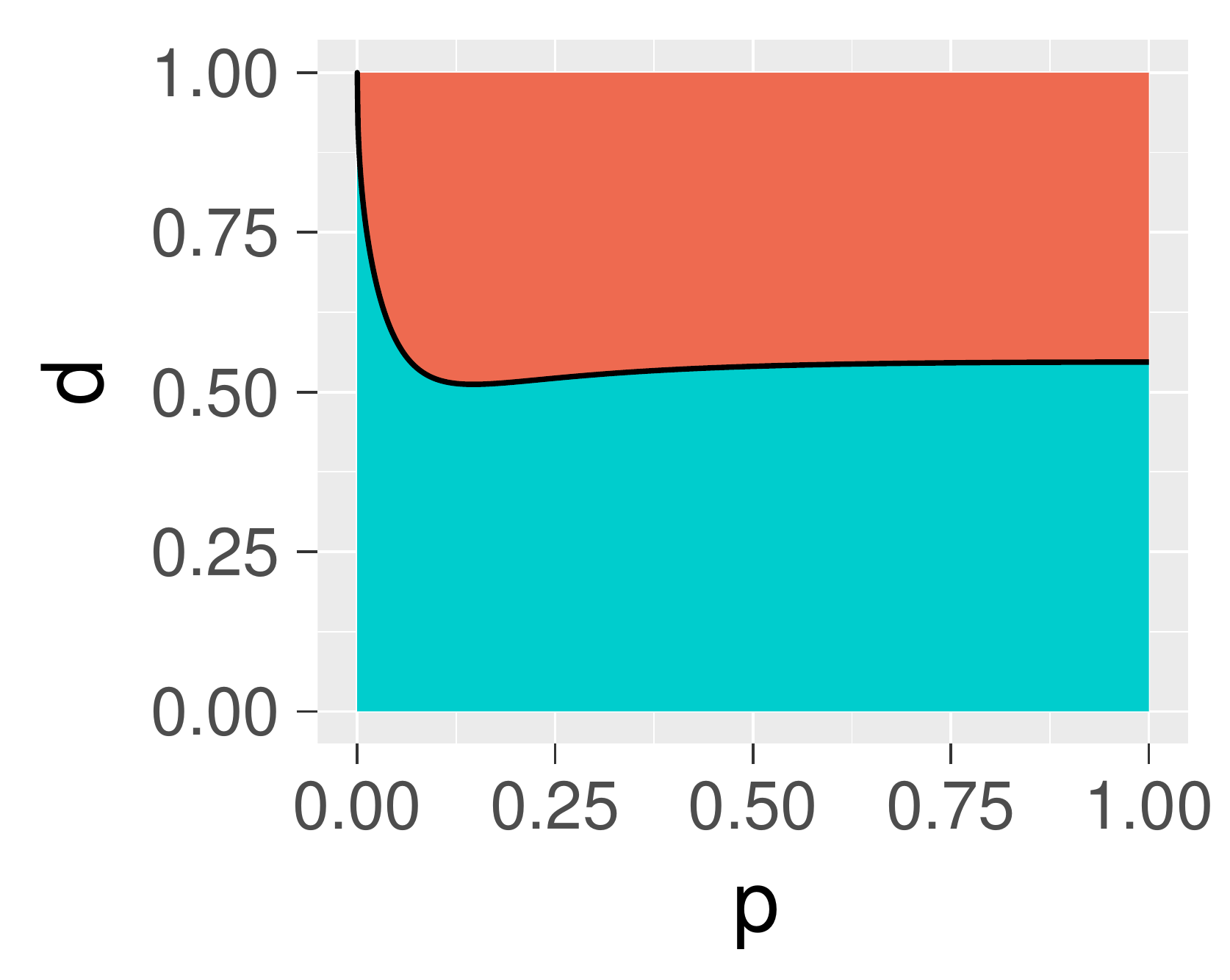}
(c)\includegraphics[width=0.45\textwidth]{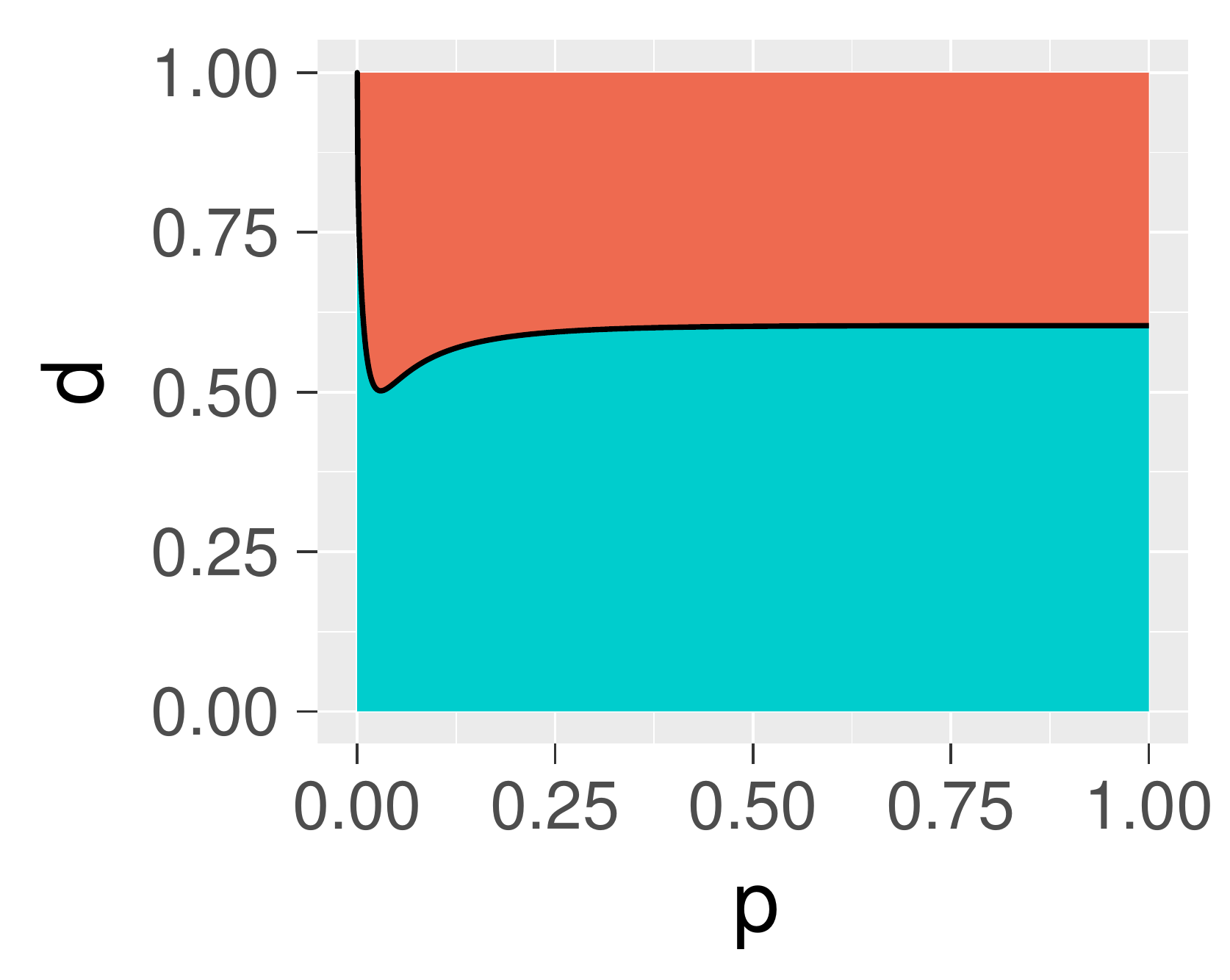}
(d)\includegraphics[width=0.45\textwidth]{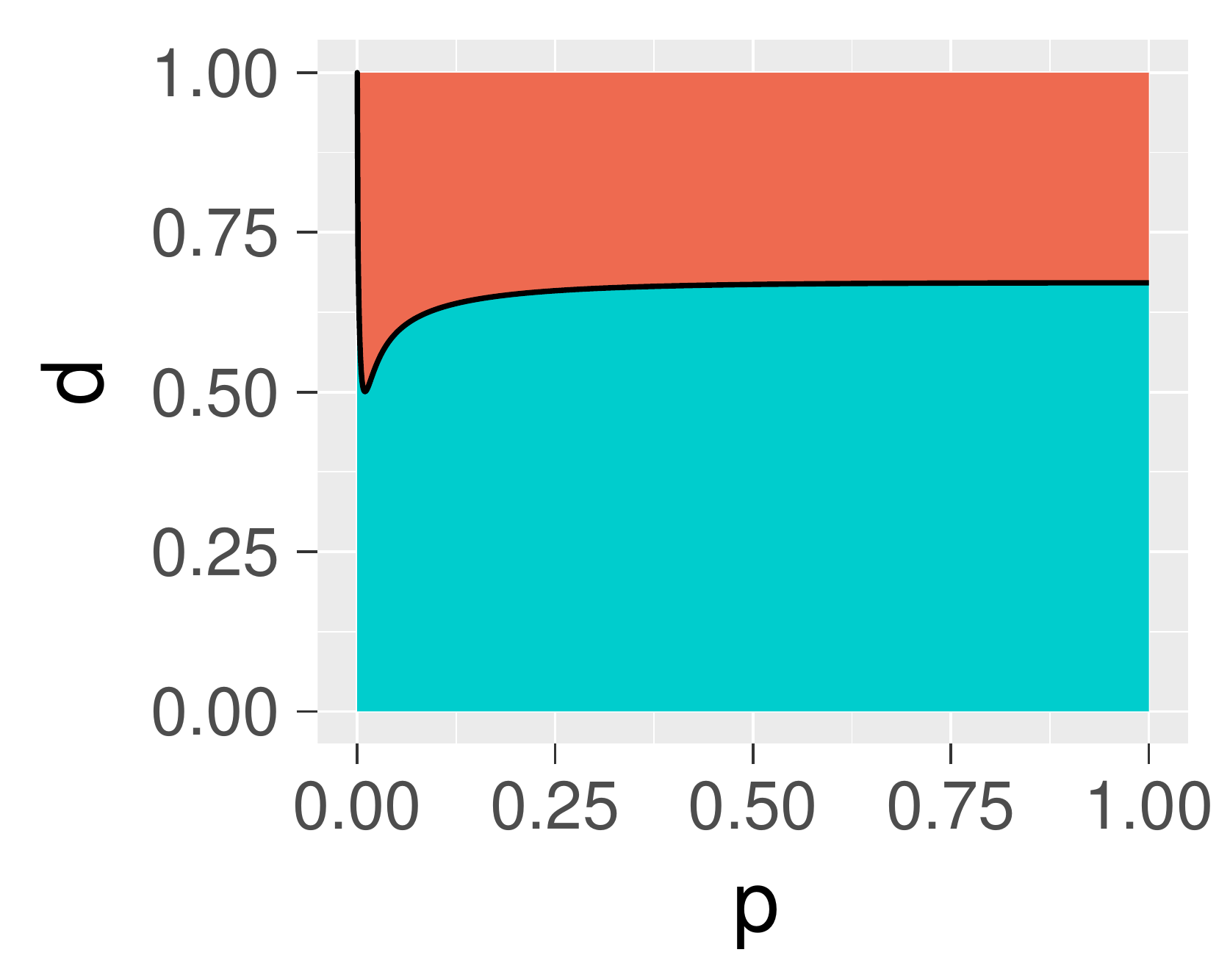}
 \caption{{\bf Phase diagram of clique splitting with excess modularity density $Q_x$ as the external influence is varied.} The values of
 clique size ratio $p$ and link density $d$ where the M phase occurs is shown in orange and where the S phase occurs is shown in blue. Results are for $\rho=\rho_{max}$ and different choices of the external influence parameter $t$: (a) t=0 (b) t=1 (c) t=5 (d) t=15.}
 \label{fig:Qxphase}
\end{figure}

\begin{figure}
    \centering
    (a)\includegraphics[width=0.45\textwidth]{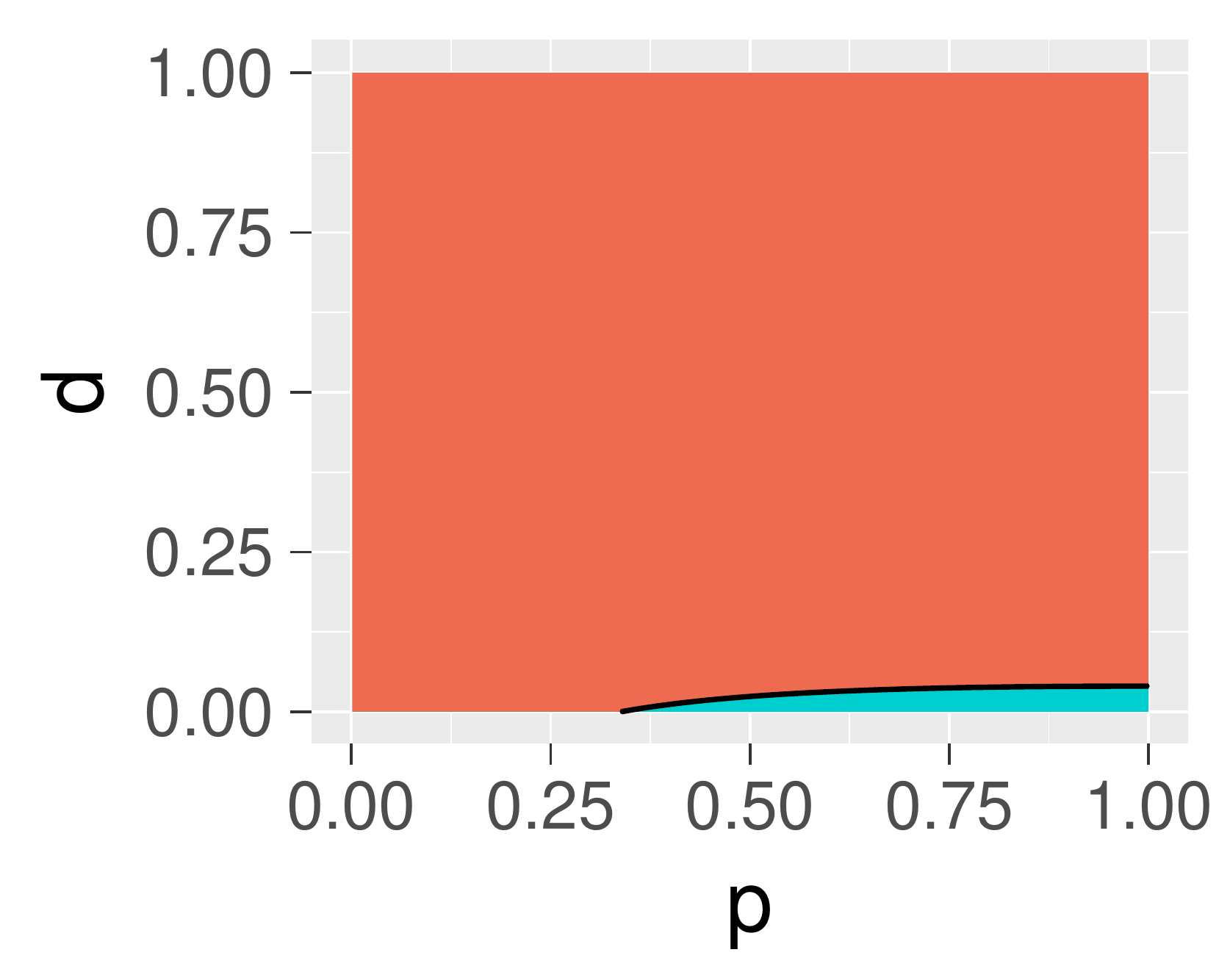}
    (b)\includegraphics[width=0.45\textwidth]{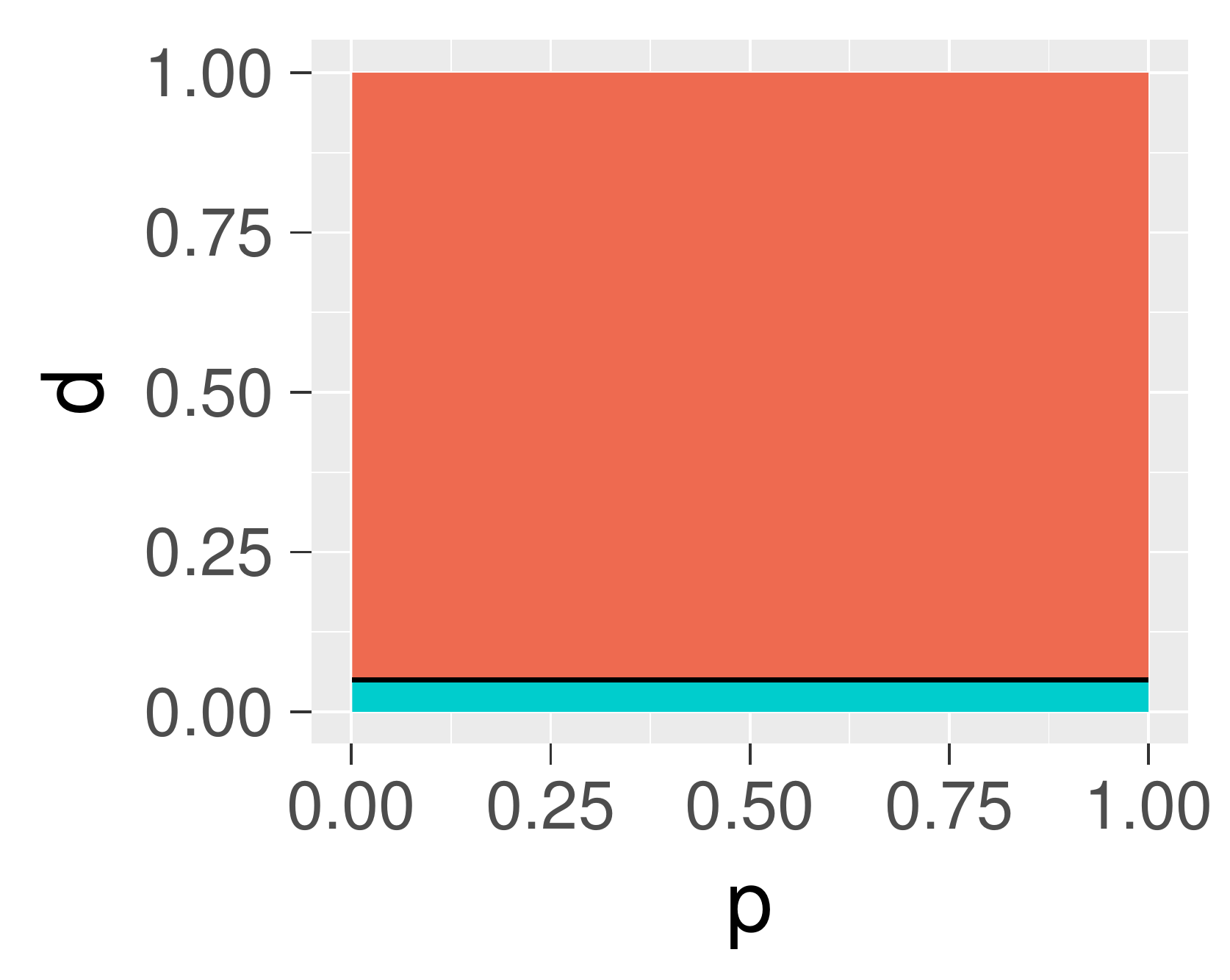}
    (c)\includegraphics[width=0.45\textwidth]{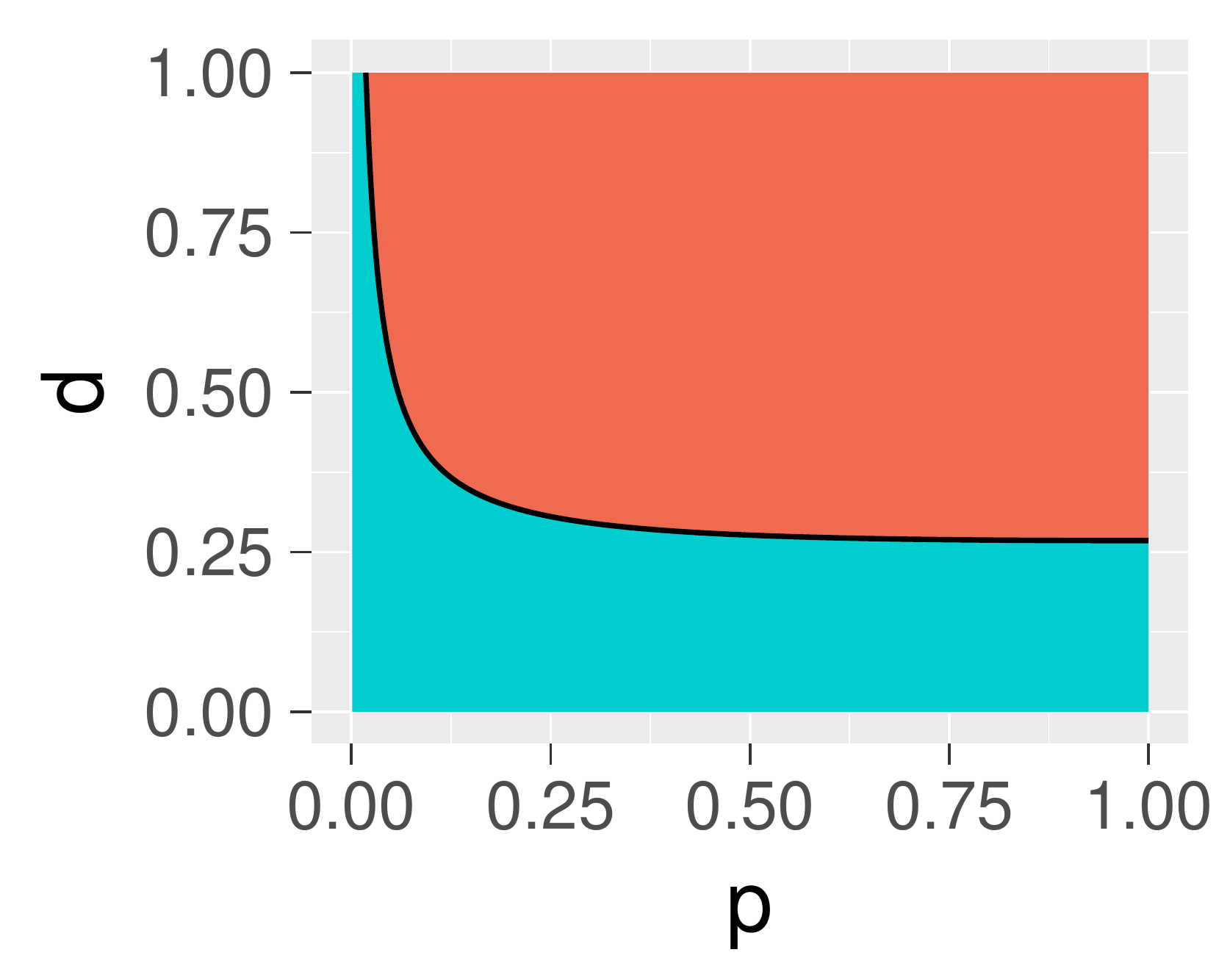}
    (d)\includegraphics[width=0.45\textwidth]{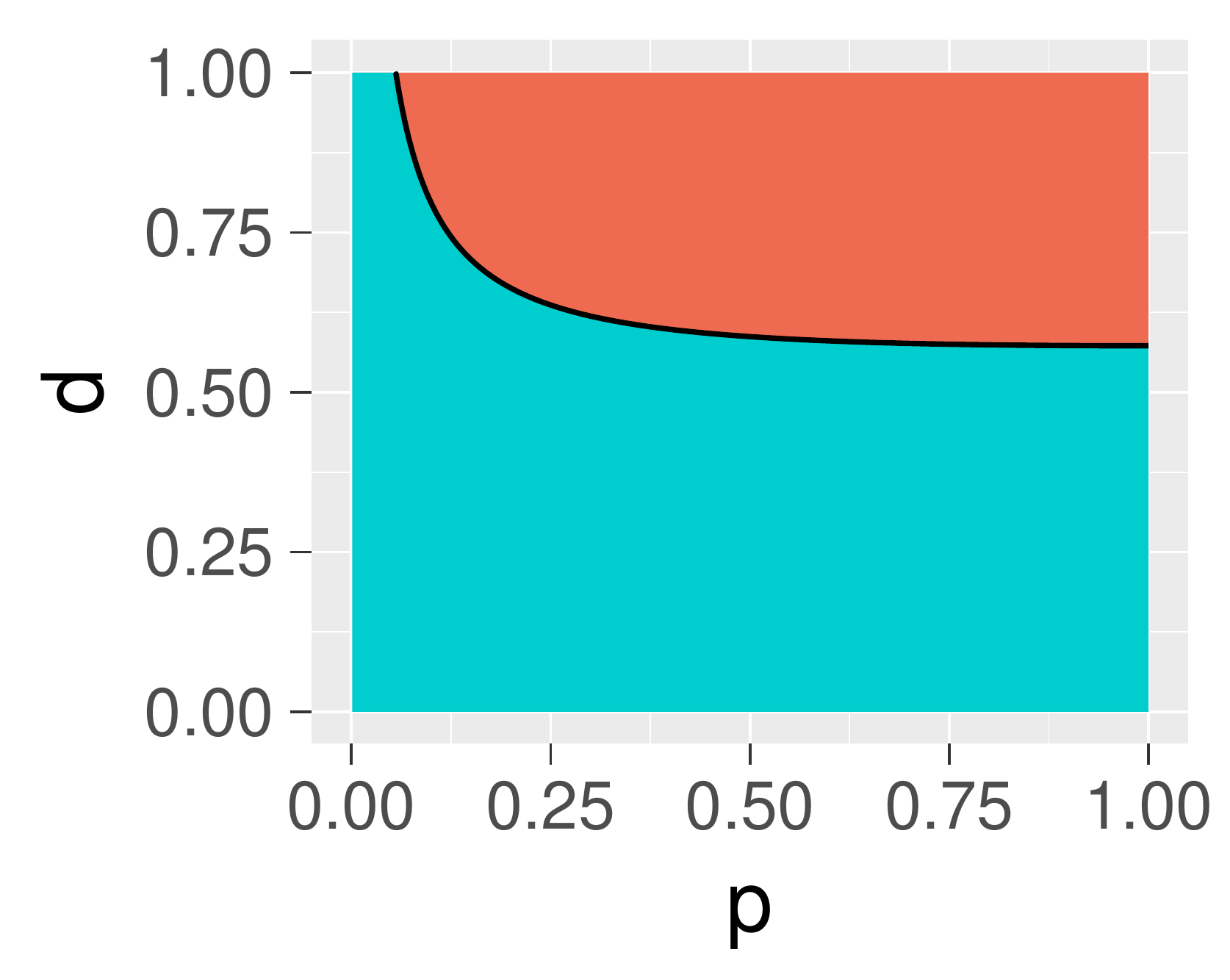}
    \caption{{\bf Phase diagram of clique splitting with $Q_{AFG}$ at fixed external influence $t$ as the parameter $s$ is varied.} The values of
 clique size ratio $p$ and link density $d$ where the M phase occurs is shown in orange and where the S phase occurs is shown in blue. Results are for $n_a=\sqrt{2m_a}$, $t=10$ and different choices of $s$: (a) $s = -\frac{m}{2N}$ (b) $s = 0$ (c)  $s = \frac{m}{2N}$ (d) $s = \frac{m}{N}$.}
    \label{fig:Qrphase}
\end{figure}

The phase diagrams of $Q$, $Q_{ds}$, $Q_w$, $Q_x(\rho=\rho_{max})$, $Q_{AFG}(n_a=\sqrt{2m_a})$ and $Q_g(\chi=1)$ are shown respectively in 
Fig.~\ref{fig:Qphase},
Fig.~\ref{fig:Qdsphase}, Fig.~\ref{fig:Qwphase}, Fig.~\ref{fig:Qxphase},
Fig~\ref{fig:Qrphase} and
Fig.~\ref{fig:Qgphase} (in the main text).
As shown in the figures, the behavior varies a lot across different metrics and the particular choice of other variables. 
In the following, we will observe some general characteristics of all phase diagrams. 
Then, we will examine each one in more detail and demonstrate that $Q_g$ performs better than other metric.\\

First, there are two phases (M (red) and S (blue) phase) in the phase diagram and as expected and M phase is above S phase implying that nearly all metrics tend to merge the two cliques when $d$ is close to 1 and to split when $d$ is close to 0. This meets the common expectation in extreme cases. But different metrics disagree when $d$ is in intermediate range.
Other variables such as $(t,\rho)$ also dictate the performance in this range.
Fig.~\ref{fig:Qphase} shows the RL problem of modularity with a much clear view.
We know that $\delta=\sqrt{t^2+1}-t$, as $t\rightarrow \infty$, we have $\delta \rightarrow 0$. This trend is also shown in the Fig.~\ref{fig:Qphase}. Therefore, given any value $d>0$, we can construct a network with large enough $t$ so that $d>\delta$, which means the two cliques, as long as they are connected, they will be merged into one community if the external component has enough links. This is the RL problem of modularity. 
However, if $d=0$, there is no RL because for any $t\geq 0$, $\delta>0$ is always true and modularity maximization would not merge two disconnected cliques. More generally it can be shown that if two subgroups of the network are disconnected, they are guaranteed to be split.\\
As shown in Fig.~\ref{fig:Qdsphase}, $Q_{ds}$ depends strongly on $p$. A different type of RL problem can be seen in the figures. If $p$ is small enough, $\delta=0$ can always be true whatever $d$ is. It means that if the sizes of two cliques are different enough, they will be merged even if $d=0$~\cite{chen2018}. It clearly violates our expectation. This problem gets alleviated as $t\rightarrow \infty$. But it always exists for arbitrary $t$.\\
For phase diagram of $Q_w$ shown in Fig.~\ref{fig:Qwphase}, the phase boundary moves down as $t\rightarrow \infty$. But it has a lower bound which means, when $d$ is small enough, the two cliques of example network  will always be split whatever other variables are. Thus, it has no extreme cases of RL as $Q$ and $Q_{ds}$. Note that M phase is reduced to a straight line $d=1$ here in Fig.~\ref{fig:Qwphase}(a) which means the extreme case of expectation is satisfied\\
As for $Q_x$, because there is one more variable $\rho$, the analysis is more complicated. As we can see from Equ.~\ref{eq:Qx} and Equ.~\ref{eq:Qds}, $Q_x(\rho \rightarrow 0)\rightarrow Q_{ds}$ which means $Q_x$ will behave the same as $Q_{ds}$ when global link density $\rho=0$. Because of the arbitrary external component, it can be easily achieved. Also we should be aware of the fact that most real-world networks are sparse thus $\rho \rightarrow 0$ is a common case where $Q_x$ will fail to solve RL as $Q_{ds}$. In Fig.~\ref{fig:Qxphase}, we show the phase diagram when $\rho$ equals to its maximum. The phase boundary, starting from $d=1$, goes down first and then rises up again. So, when $t\rightarrow \infty$, the two cliques will always be split as long as $d<1$. But this requires both $\rho,t$ are very large which is uncommon for most real-world networks. \\

We use the AFG method~\cite{arenas2008} on the benchmark network, which attempts to solve the RL problem by assigning a self loop of weight $s$ to each node. This method allows one to explores communities at different resolution densities by controlling $s$. Using $Q_g$, this is achieved by controlling $\chi$ so the two methods are similar in spirit. However, irrespective of the choice of $s$, the metric $Q_{AFG}$ will behave like modularity $Q$ and fail to resolve clusters if $n_a$ in the benchmark network is sufficiently large. To avoid that, we show the phase diagram (Fig.~\ref{fig:Qrphase}) for $n_a = \sqrt{2m_a}$, which is the smallest possible $n_a$ for a fixed $m_a$ (in the large $m_a$ limit) and perhaps the best case scenario for $Q_{AFG}$.
Moreover, if a specific resolution density is desired then $s$ must be selected according to the network size, unlike $Q_g$, which has the lower bound that is independent of the network size. Even if $s$ is chosen according to the network size, the phase diagram in Fig.~\ref{fig:Qrphase}~(a), (c), (d) shows that the metric $Q_{AFG}$ will fail when the two cliques are somewhat different in size (small $p$). When $p$ is small and $s \ne 0$, it either merges two disconnected cliques (Fig.~\ref{fig:Qrphase}~(a)), or splits a larger clique formed by clique 1 and clique 2 (Fig.~\ref{fig:Qrphase}~(c) and (d)). When $s = 0$ (Fig.~\ref{fig:Qrphase}~(b)), the metric $Q_{AFG}$ is the same as modularity $Q$ and it will have the same problems as outlined before.
The phase diagram also shows that for a non-zero value of $s$, the resolution density varies a lot as a function of $p$. This implies that merging or splitting the two cliques is heavily influenced by their relative sizes. Thus, in a network with a wide range of community sizes, this method will be biased either towards merging well separated communities or splitting well connected communities, an observation also made in~\cite{lancichinetti2011}.

\pagebreak

\end{document}